\documentclass[twocolumn]{aastex61}
\pdfoutput=1 
\usepackage{amsmath,amstext}
\usepackage[T1]{fontenc}
\usepackage{apjfonts} 
\usepackage[figure,figure*]{hypcap}

\newcommand{\attw}{{ATLAS12}}
\newcommand\halpha{H$\alpha$} 
\newcommand\vsini{$v$\,sin\,$i_\star$}   
\newcommand\vmic{$v_{\rm mic}$}
\newcommand\vmac{$v_{\rm mac}$}
\newcommand\teff{$T_{\rm eff}$}
\newcommand\logg{log\,{\it g}}
\newcommand\met{[Fe/H]}

\shorttitle{EPIC~228732031\lowercase{b}}
\shortauthors{Dai et al}

\begin{document}

\title{The discovery and mass measurement of a new ultra-short-period planet: EPIC~228732031b}

\author{Fei Dai}
\affiliation{Department of Physics and Kavli Institute for Astrophysics and Space Research, Massachusetts Institute of Technology, Cambridge, MA, 02139, USA}
\affiliation{Department of Astrophysical Sciences, Princeton University, 4 Ivy Lane, Princeton, NJ, 08544, USA}
\email{fd284@mit.edu}

\author{Joshua N. Winn}
\affiliation{Department of Astrophysical Sciences, Princeton University, 4 Ivy Lane, Princeton, NJ, 08544, USA}

\author{Davide Gandolfi}
\affiliation{Dipartimento di Fisica, Universit\`a di Torino, via P. Giuria 1, 10125 Torino, Italy}

\author{Sharon X. Wang}
\affiliation{Department of Terrestrial Magnetism, Carnegie Institution for Science, 5241 Broad Branch Road, NW, Washington DC, 20015-1305, USA}

\author{Johanna K. Teske}
\affiliation{The Observatories of the Carnegie Institution for Science, 813 Santa Barbara Street, Pasadena, CA 91101, USA}
\affiliation{Carnegie Origins Fellow, jointly appointed by Carnegie DTM and Observatories}

\author{Jennifer Burt}
\affiliation{Department of Physics and Kavli Institute for Astrophysics and Space Research, Massachusetts Institute of Technology, Cambridge, MA, 02139, USA}

\author{Simon Albrecht}
\affiliation{Stellar Astrophysics Centre, Department of Physics and Astronomy, Aarhus University, Ny Munkegade 120, DK-8000 Aarhus C, Denmark}

\author{Oscar Barrag\'an}
\affiliation{Dipartimento di Fisica, Universit\`a di Torino, via P. Giuria 1, 10125 Torino, Italy}

\author{William D. Cochran}
\affiliation{Department of Astronomy and McDonald Observatory, University of Texas at Austin, 2515 Speedway, Stop C1400, Austin, TX 78712, USA}

\author{Michael Endl}
\affiliation{Department of Astronomy and McDonald Observatory, University of Texas at Austin, 2515 Speedway, Stop C1400, Austin, TX 78712, USA}

\author{Malcolm Fridlund}
\affiliation{Leiden Observatory, University of Leiden, PO Box 9513, 2300 RA, Leiden, The Netherlands}
\affiliation{Department of Space, Earth and Environment, Chalmers University of Technology, Onsala Space Observatory SE - 439 92 Onsala, Sweden}

\author{Artie P. Hatzes}
\affiliation{Th\"uringer Landessternwarte Tautenburg, Sternwarte 5, D-07778 Tautenberg, Germany}

\author{Teruyuki Hirano}
\affiliation{Department of Earth and Planetary Sciences, Tokyo Institute of Technology, 2-12-1 Ookayama, Meguro-ku, Tokio 152-8551, Japan}

\author{Lea A. Hirsch}
\affiliation{University of California at Berkeley, Berkeley, CA 94720, USA}

\author{Marshall C. Johnson}
\affiliation{Department of Astronomy, The Ohio State University, 140 West 18th Ave., Columbus, OH 43210, USA}

\author{Anders Bo Justesen}
\affiliation{Stellar Astrophysics Centre, Department of Physics and Astronomy, Aarhus University, Ny Munkegade 120, DK-8000 Aarhus C, Denmark}

\author{John Livingston}
\affiliation{Department of Astronomy, The University of Tokyo, 7-3-1 Hongo, Bunkyo-ku, Tokyo 113-0033, Japan}

\author{Carina M. Persson}
\affiliation{Department of Space, Earth and Environment, Chalmers University of Technology, Onsala Space Observatory SE - 439 92 Onsala, Sweden}

\author{Jorge Prieto-Arranz}
\affiliation{Instituto de Astrof\'isica de Canarias, C/V\'ia L\'actea s/n, 38205 La Laguna, Spain}
\affiliation{Departamento de Astrof\'isica, Universidad de La Laguna, 38206 La Laguna, Spain}

\author{Andrew Vanderburg}
\affiliation{Harvard-Smithsonian Center for Astrophysics, 60 Garden Street, Cambridge, MA 02138, USA}

\author{Roi Alonso}
\affiliation{Instituto de Astrof\'isica de Canarias, C/V\'ia L\'actea s/n, 38205 La Laguna, Spain}
\affiliation{Departamento de Astrof\'isica, Universidad de La Laguna, 38206 La Laguna, Spain}

\author{Giuliano Antoniciello}
\affiliation{Dipartimento di Fisica, Universit\'a di Torino, via P. Giuria 1, 10125 Torino, Italy}

\author{Pamela Arriagada}
\affiliation{Department of Terrestrial Magnetism, Carnegie Institution for Science, 5241 Broad Branch Road, NW, Washington DC, 20015-1305, USA}

\author{R.P. Butler}
\affiliation{Department of Terrestrial Magnetism, Carnegie Institution for Science, 5241 Broad Branch Road, NW, Washington DC, 20015-1305, USA}

\author{Juan Cabrera}
\affiliation{Institute of Planetary Research, German Aerospace Center, Rutherfordstrasse 2, D-12489 Berlin, Germany}

\author{Jeffrey D. Crane}
\affiliation{The Observatories of the Carnegie Institution for Science, 813 Santa Barbara Street, Pasadena, CA 91101, USA}

\author{Felice Cusano}
\affiliation{INAF - Osservatorio Astronomico di Bologna, Via Ranzani, 1, 20127, Bologna, Italy}

\author{Szil\'ard Csizmadia}
\affiliation{Institute of Planetary Research, German Aerospace Center, Rutherfordstrasse 2, D-12489 Berlin, Germany}

\author{Hans Deeg}
\affiliation{Instituto de Astrof\'isica de Canarias, C/V\'ia L\'actea s/n, 38205 La Laguna, Spain}
\affiliation{Departamento de Astrof\'isica, Universidad de La Laguna, 38206 La Laguna, Spain}

\author{Sergio B. Dieterich}
\affiliation{Department of Terrestrial Magnetism, Carnegie Institution for Science, 5241 Broad Branch Road, NW, Washington DC, 20015-1305, USA}
\affiliation{NSF Astronomy and Astrophysics Postdoctoral Fellow}

\author{Philipp Eigm\"uller}
\affiliation{Institute of Planetary Research, German Aerospace Center, Rutherfordstrasse 2, D-12489 Berlin, Germany}

\author{Anders Erikson}
\affiliation{Institute of Planetary Research, German Aerospace Center, Rutherfordstrasse 2, D-12489 Berlin, Germany}

\author{Mark E. Everett}
\affiliation{National Optical Astronomy Observatory, 950 N. Cherry Ave., Tucson, AZ 85719, USA}

\author{Akihiko Fukui}
\affiliation{Okayama Astrophysical Observatory, National Astronomical Observatory of Japan, Asakuchi, 719-0232 Okayama, Japan}

\author{Sascha Grziwa}
\affiliation{Rheinisches Institut f\"ur Umweltforschung an der Universit\"at zu K\"oln, Aachener Strasse 209, 50931 K\"oln, Germany}

\author{Eike W. Guenther}
\affiliation{Th\"uringer Landessternwarte Tautenburg, Sternwarte 5, D-07778 Tautenberg, Germany}

\author{Gregory W. Henry}
\affiliation{Center of Excellence in Information Systems, Tennessee State University, Nashville, TN 37209}

\author{Steve B. Howell}
\affiliation{NASA Ames Research Center, Moffett Field, CA 94035, USA}

\author{John Asher Johnson}
\affiliation{Harvard-Smithsonian Center for Astrophysics, 60 Garden Street, Cambridge, MA 02138, USA}

\author{Judith Korth}
\affiliation{Rheinisches Institut f\"ur Umweltforschung an der Universit\"at zu K\"oln, Aachener Strasse 209, 50931 K\"oln, Germany}

\author{Masayuki Kuzuhara}
\affiliation{Astrobiology Center, NINS, 2-21-1 Osawa, Mitaka, Tokyo 181-8588, Japan}
\affiliation{National Astronomical Observatory of Japan, NINS, 2-21-1 Osawa, Mitaka, Tokyo 181-8588, Japan}

\author{Norio Narita}
\affiliation{Department of Astronomy, The University of Tokyo, 7-3-1 Hongo, Bunkyo-ku, Tokyo 113-0033, Japan}
\affiliation{Astrobiology Center, NINS, 2-21-1 Osawa, Mitaka, Tokyo 181-8588, Japan}
\affiliation{National Astronomical Observatory of Japan, NINS, 2-21-1 Osawa, Mitaka, Tokyo 181-8588, Japan}

\author{David Nespral}
\affiliation{Instituto de Astrof\'isica de Canarias, C/V\'ia L\'actea s/n, 38205 La Laguna, Spain}
\affiliation{Departamento de Astrof\'isica, Universidad de La Laguna, 38206 La Laguna, Spain}

\author{Grzegorz Nowak}
\affiliation{Instituto de Astrof\'isica de Canarias, C/V\'ia L\'actea s/n, 38205 La Laguna, Spain}
\affiliation{Departamento de Astrof\'isica, Universidad de La Laguna, 38206 La Laguna, Spain}

\author{Enric Palle}
\affiliation{Instituto de Astrof\'isica de Canarias, C/V\'ia L\'actea s/n, 38205 La Laguna, Spain}
\affiliation{Departamento de Astrof\'isica, Universidad de La Laguna, 38206 La Laguna, Spain}

\author{Martin P\"atzold}
\affiliation{Rheinisches Institut f\"ur Umweltforschung an der Universit\"at zu K\"oln, Aachener Strasse 209, 50931 K\"oln, Germany}

\author{Heike Rauer}
\affiliation{Institute of Planetary Research, German Aerospace Center, Rutherfordstrasse 2, D-12489 Berlin, Germany}
\affiliation{Center for Astronomy and Astrophysics, TU Berlin, Hardenbergstr. 36, D-10623 Berlin, Germany}

\author{Pilar~Monta\~n\'es Rodr\'iguez}
\affiliation{Instituto de Astrof\'isica de Canarias, C/V\'ia L\'actea s/n, 38205 La Laguna, Spain}
\affiliation{Departamento de Astrof\'isica, Universidad de La Laguna, 38206 La Laguna, Spain}

\author{Stephen A. Shectman}
\affiliation{The Observatories of the Carnegie Institution for Science, 813 Santa Barbara Street, Pasadena, CA 91101, USA}

\author{Alexis M.S. Smith}
\affiliation{Institute of Planetary Research, German Aerospace Center, Rutherfordstrasse 2, D-12489 Berlin, Germany}

\author{Ian B. Thompson}
\affiliation{The Observatories of the Carnegie Institution for Science, 813 Santa Barbara Street, Pasadena, CA 91101, USA}

\author{Vincent Van Eylen}
\affiliation{Leiden Observatory, University of Leiden, PO Box 9513, 2300 RA, Leiden, The Netherlands}

\author{Michael W. Williamson}
\affiliation{Center of Excellence in Information Systems, Tennessee State University, Nashville, TN 37209}

\author{Robert A. Wittenmyer}
\affiliation{University of Southern Queensland, Computational Science and Engineering Research Centre, Toowoomba QLD Australia}

\begin{abstract}
We report the discovery of a new ultra-short-period planet and summarize the properties of all such planets
for which the mass and radius have been measured.
The new planet, EPIC~228732031b, was discovered in {\it K2} Campaign 10. It has a radius of 1.81$^{+0.16}_{-0.12}~R_{\oplus}$ and orbits a G dwarf with a period of 8.9 hours.
Radial velocities obtained with Magellan/PFS and TNG/HARPS-N show evidence for stellar activity along with orbital motion. 
We determined the planetary mass using two different methods: (1) the "floating chunk offset" method, based only on changes
in velocity observed on the same night; and (2) a Gaussian process regression
based on both the radial-velocity and photometric time series.
The results are consistent and lead to a mass measurement of $6.5 \pm 1.6~M_{\oplus}$, and a mean density of
$6.0^{+3.0}_{-2.7}$~g~cm$^{-3}$.
\end{abstract}


\section{Introduction}

The ultra-short-period (USP) planets, with orbital periods
shorter than one day, are usually smaller than about
2~$R_\oplus$. A well-studied example is Kepler-78b, a roughly Earth-sized planet with an 8.5-hour orbit around a solar-type star \citep{Sanchis-Ojedak78,Howard2013,Pepe2013}. Using {\it Kepler} data, \citet{Sanchis-Ojeda_usp} presented a sample of about 100 transiting USP planets. They found their occurrence rate to be about 0.5\% around G-type dwarf stars, with higher rates for
KM stars and a lower rate for F stars. They also noted that many if not all of the USP planets have wider-orbiting planetary companions. It has been postulated that USP planets
were once somewhat larger planets that lost their gaseous envelopes \citep{Sanchis-Ojeda_usp,Lopez2016,Lundkvist2016,Winn2017},
perhaps after undergoing tidal orbital decay \citep{LeeChiang2017}.

\citet{Fulton2017} reported evidence supporting the notion that planets with a hydrogen-helium (H/He) envelope can undergo photoevaporation, shrinking their size from 2-3~$R_{\oplus}$ to 1.5~$R_\oplus$ or smaller. Specifically, they found the size distribution of close-in ($P_{\text{orb}}$ < 100 days) {\it Kepler} planets to be bimodal, with a dip in occurrence between 1.5-2~$R_\oplus$.
\citet{Owen2013} and \citet{Lopez2014} had predicted such a dip as a consequence of photoevaporation.
\citet{Owen2017} further demonstrated that the observed radius distribution can be reproduced by a model in which photoevaporation is applied to a single population of super-Earths with gaseous envelopes.

Thus, the USP planets are interesting for further tests and refinements of the photoevaporation theory. They are typically bathed in stellar radiation with a flux $>$10$^3$ higher than the Earth's insolation, where theory predicts they should be rocky cores entirely stripped of H/He gas. By studying their distribution in mass, radius, and orbital distance, we may learn about the primordial population of rocky cores and the conditions in which they formed. So far, though, masses have been measured for only a handful of USP planets. The main limitation has been the relative faintness of their host stars, which are drawn mainly from the {\it Kepler} survey.

In this paper, we present the discovery and Doppler mass measurement of another USP planet,
EPIC~228732031b. The host star is a G-type dwarf with $V=12.1$
that was observed in {\it K2} Campaign 10.
This paper is organized as follows. Section 2 presents time-series photometry of EPIC~228732031, both space-based and ground-based. Section 3 describes our radial velocity (RV) observations.
Section 4 presents high-angular-resolution images of the field surrounding EPIC~228732031 and the resultant constraints on any nearby companions. Section 5 is concerned with the stellar parameters of EPIC~228732031, as determined by spectroscopic analysis and stellar-evolutionary models.
Section 6 presents an analysis of the time-series photometry, including the transit detection, light-curve modeling, and measurement of the stellar rotation period.
Section 7 describes the two different methods we employed to analyze the RV data.
Section 8 summarizes the properties of all the known USP planets for which mass and radius have
been measured.

\section{Photometric Observations}

\subsection{{\it K2}}

EPIC~228732031 was observed by the {\it Kepler} spacecraft from July 6 to September 20, 2016, during {\it K2} Campaign 10. According to the {\it K2} Data Release Notes\footnote{keplerscience.arc.nasa.gov/k2-data-release-notes.htm.}, there was a 3.5-pixel pointing error during the first 6 days of Campaign 10, degrading the data quality. We discarded the data obtained during this period. Later in Campaign 10, the loss of Module 4 resulted in a 14-day gap in data collection. Therefore, the light curves consist of an initial interval of about 6 days, followed by the 14-day data gap, and another continuous interval of about 50 days.

\begin{figure*}
\begin{center}
\includegraphics[width = 2.\columnwidth]{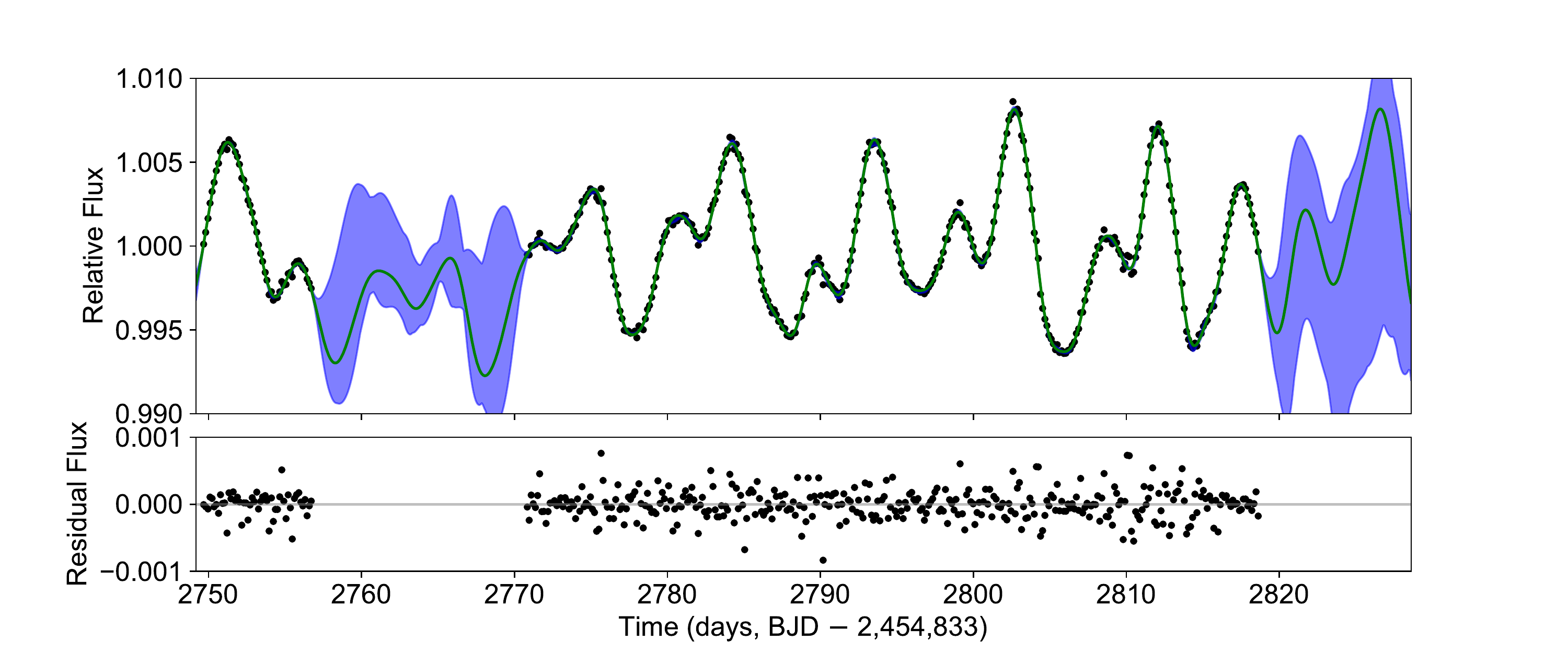}
\caption{{\it K2} light curve of EPIC~228732031, after removing the transits of planet b. The black circles are binned fluxes. The light curve shows a rotational modulation with a period of 9.4 days and an amplitude of about 0.5\%. The green curve shows the Gaussian Process regression of the {\it K2} light curve with a quasi-periodic kernel (Section \ref{sec:gp}). The blue shaded region is the 1$\sigma$ confidence interval of the Gaussian Process.}
\label{fig:K2_lc}
\end{center}
\end{figure*}

To produce the light curve, we downloaded the target pixel files from the Mikulski Archive for Space Telescopes.\footnote{https://archive.stsci.edu/k2.} We then attempted to reduce
the well-known apparent brightness fluctuations associated with the rolling motion of the spacecraft,
adopting an approach similar to that described by  \citet{VJ2014}. For each image, we laid down a circular aperture around the brightest pixel, and fitted a 2-d Gaussian function to the intensity distribution. We then fitted a piecewise linear function between the observed flux variation and the central coordinates of the Gaussian function. Figure \ref{fig:K2_lc} shows the detrended {\it K2} light curve of EPIC~228732031.

\subsection{AIT}

Since the {\it K2} light curve showed signs of stellar activity (as discussed in Section~\ref{sec:photometric}), we scheduled ground-based photometric observations of EPIC~228732031, overlapping in time with our RV follow-up campaign. Our hope was that the observed photometric variability could be used to disentangle the effects of stellar activity and orbital motion.

We observed EPIC~228732031 nightly with the Tennessee State University Celestron 14-inch (C14) Automated Imaging Telescope (AIT) located at Fairborn Observatory, Arizona \citep[see, e.g.,][]{h1999}. The observations were made in the Cousins $R$ bandpass.  Each nightly observation consisted of 4-10 consecutive exposures of the field centered on EPIC~228732031. The nightly observations were corrected for bias, flat-fielding, and differential atmospheric extinction. The individual reduced frames were co-added and aperture photometry was carried out on each co-added frame. We performed ensemble differential photometry, i.e., the mean instrumental magnitude of the six comparison stars was subtracted from the instrumental magnitude of EPIC~228732031. Table~\ref{tab:AIT} provides the 149 observations that were collected between March 15 and May 2, 2017.

\subsection{Swope}

EPIC~228732031 was monitored for photometric variability in the Bessel $V$
band from March 21 to April 1, 2017
using the Henrietta Swope 1m telescope at Las Campanas
Observatory.
Exposures of 25s were taken consecutively for 2 hours
at the beginning and the end of each night if weather permits.
The field of view of the images was $28\arcsec \times 28\arcsec$.
Initially we selected 59 stars as candidate
reference stars for differential aperture photometry.
The differential light curve of each star was obtained
by dividing the flux of each star by the sum of the fluxes
of all the reference stars.
The candidate reference stars were then ranked in order of increasing
variability.  Light curves of EPIC~228732031 were calculated
using successively larger numbers of these rank-ordered
reference stars.  The noise level was found to be minimized
when the 16 top-ranked candidate reference stars were used; this collection
of stars was adopted to produce the final light curve of EPIC~228732031.
Since we are interested in the long-term variability, we binned the 25s exposures taken within each 2-hour window. The relative flux measurements
and uncertainties are provided in Table~\ref{tab:Swope}.
\newline
\newline

\section{Radial Velocity Observations}

\subsection{HARPS-N}

Between January 29 and April 1, 2017 (UT), we collected 41 spectra of EPIC~228732031 using the HARPS-N spectrograph \citep[R\,$\approx$\,115000;][]{Cosentino2012} mounted on the 3.58m Telescopio Nazionale Galileo (TNG) of Roque de los Muchachos Observatory, in La Palma. The observations were carried out as part of the observing programs A33TAC\_15 and A33TAC\_11. We set the exposure time to 1800-2400 sec and obtained multiple spectra per night. The data were reduced using the HARPS-N off-line pipeline. RVs were extracted by cross-correlating the extracted \'echelle spectra with a K0 numerical mask \citep{Pepe2002}. Table~\ref{HARPS-N} reports the
time of observation, RV, internally-estimated measurement uncertainty, full-width half maximum (FWHM) and bisector span (BIS) of the cross-correlation function (CCF), the Ca\,{\sc ii} H\,\&\,K chromospheric activity index (log\,$R^\prime_\mathrm{HK}$), the corresponding uncertainties ($\Delta$ log\,$R^\prime_\mathrm{HK}$) and the signal-to-noise ratio (SNR) per pixel at 5500~\AA. 

\subsection{Planet Finder Spectrograph}

We also observed EPIC~228732031 between March 16 and April 5 (UT), 2017, with the Carnegie Planet Finder Spectrograph \citep[PFS, R\,$\approx$\,76000, ][]{Crane2010} on the 6.5m Magellan/Clay Telescope at Las Campanas Observatory, Chile. We adopted a similar strategy of obtaining multiple observations during each night. We took two consecutive frames for each visit, and attempted 3-5 visits per night. We obtained a total of 32 spectra in 6 nights. The detector was read out in the 2$\times$2 binned mode. The exposure time was set to 1200 sec. We obtained a separate spectrum with higher resolution and SNR, without the iodine cell, to use as a template spectrum. The RVs were determined with the technique of \citet{Butler}. The internal measurement uncertainties were estimated from the scatter in the results to fitting individual 2~\AA~sections of the spectrum. The uncertainties ranged from 3-6~m~s$^{-1}$. Table~\ref{PFS} gives the time of observation, RV, internally-estimated measurement uncertainty, and the Ca\,{\sc ii} H\,\&\,K chromospheric activity indicator $S_{\text{HK}}$.

\section{High Angular Resolution Imaging}

\subsection{Speckle Imaging}

\begin{figure}
\centering
\includegraphics[width = 1.\columnwidth]{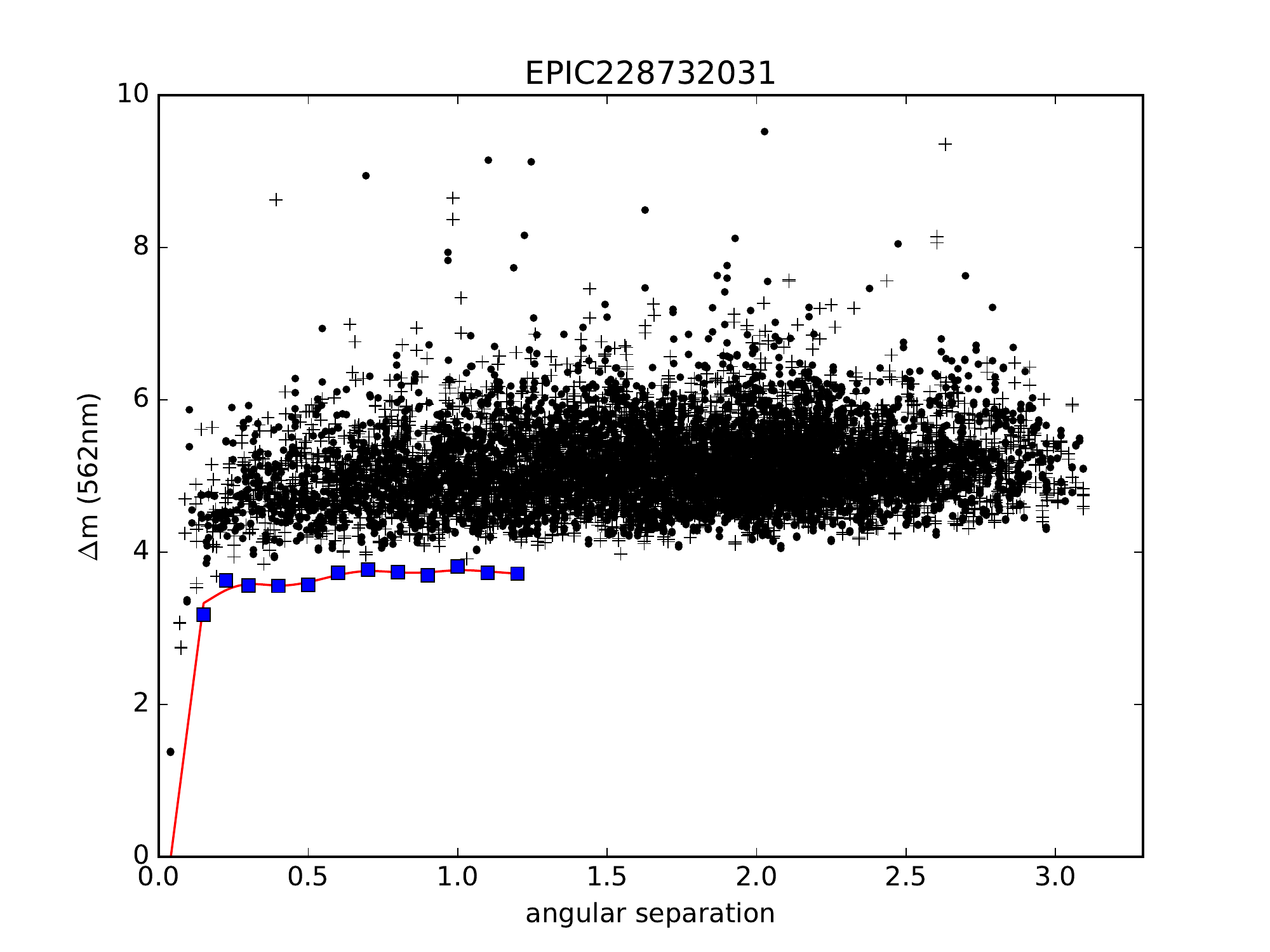}
\includegraphics[width = 1.\columnwidth]{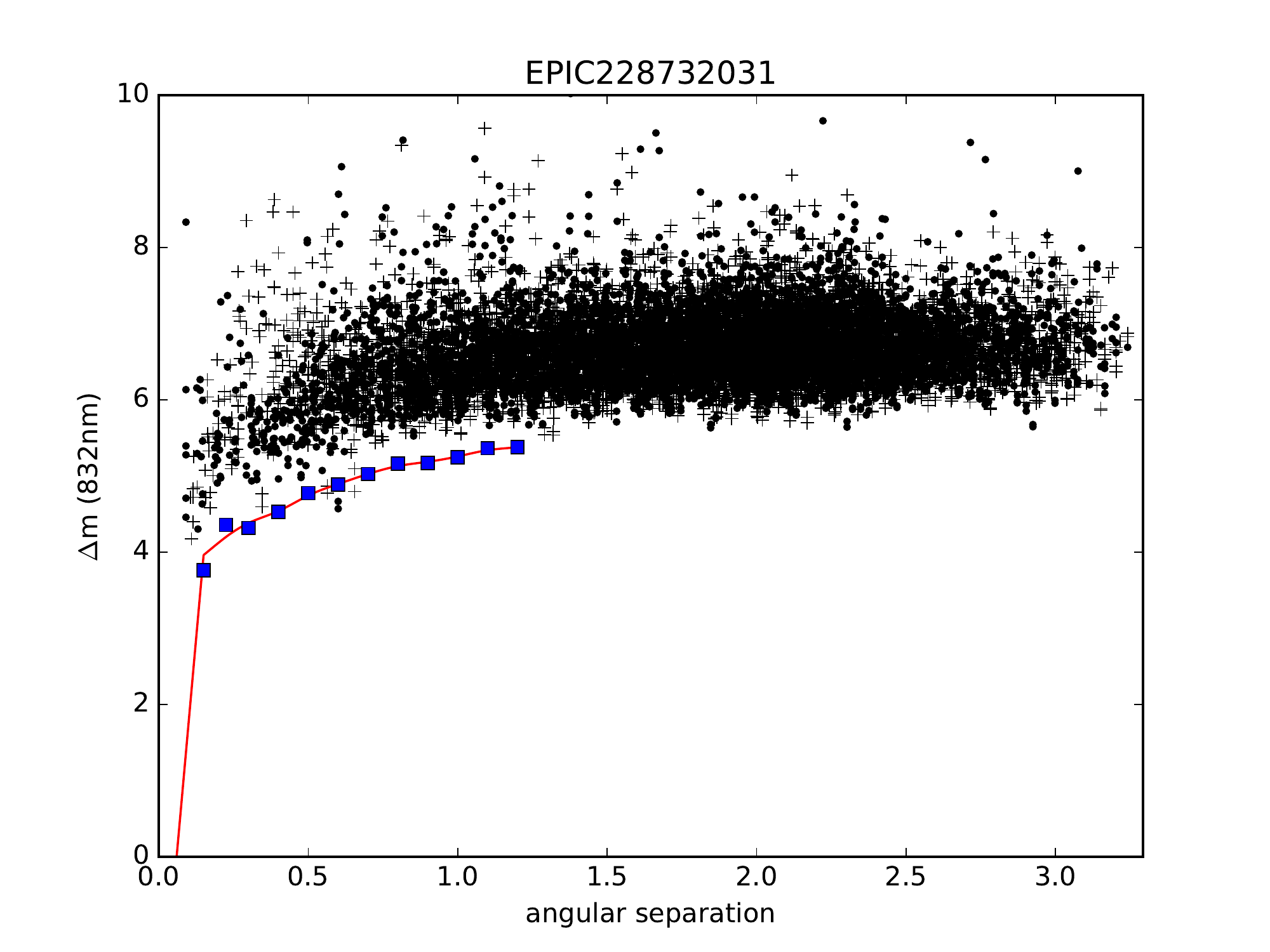}

\caption{The $5\sigma$ contrast curve based on the speckle images obtained with WIYN/NESSI. The upper panel shows a ''blue'' band centered at 562\,nm with a width of 44\,nm, and the lower panel shows a ``red'' band centered at 832\,nm with a width of 40\,nm. The blue squares are 5$\sigma$ sensitivity limits as a function of angular separation. No secondary sources were detected in the reconstructed images. The data points represent local extrema measured in the background sky of our reconstructed speckle image. Plus signs are local maxima and dots are local minima. The blue squares show the $5\sigma$ background sensitivity limit and the smooth curve is the spline fit.}
\label{speckle}
\end{figure}

On the night of April 5 (UT), 2017, we observed EPIC~228732031 with the NASA Exoplanet Star and Speckle Imager (NESSI), as part of an approved NOAO observing program (P.I.\ Livingston, proposal ID 2017A-0377). 
NESSI is a new instrument for the 3.5m WIYN Telescope \citep[Scott et al, in prep;][]{Scott2016}. It
uses high-speed electron-multiplying CCDs to capture sequences of 40~ms exposures
simultaneously in two bands: a ``blue'' band centered at 562~nm with a width of 44~nm, and a ``red'' band centered at 832~nm with a width of 40~nm. We also observed nearby point-source calibrator stars close in time. We conducted all observations in the two bands simultaneously. Using the point-source calibrator images, we reconstructed 256$\times$256 pixel images in each band, corresponding to 4.6$\arcsec$$\times$4.6$\arcsec$. No secondary sources were detected in the reconstructed images. We could exclude companions
brighter than 3\% and 1\% of the target star respectively in the blue and red band at separation
of 1$\arcsec$. We measured the background sensitivity of the reconstructed images using a series of concentric annuli centered on the target star, resulting in 5$\sigma$ sensitivity limits as a function of angular separation. The resultant contrast curves are plotted in Figure~\ref{speckle}.

\subsection{Adaptive Optics}

On the night of May 23 (UT), 2017, we performed adaptive optics (AO) imaging of EPIC~228732031
with the Infrared Camera and Spectrograph \citep[IRCS:][]{Kobayashi2000} mounted 
on the 8.2m Subaru Telescope. To search for nearby faint companions around EPIC~228732031
we obtained lightly saturated frames using the $H$-band filter with individual exposure times 
of 10s. We co-added the exposures in groups of three.
The observations were performed in the high-resolution mode (1 pixel = 20.6 mas)
using five-point dithering to minimize the impact of bad and hot pixels. We repeated 
the integration sequence for a total exposure time of 300~sec.
For absolute 
flux calibration, we also obtained unsaturated frames in which the individual exposure time was 
set to 0.412~sec, and co-added three exposures.

\begin{figure}
\centering\includegraphics[width = 1.\columnwidth]{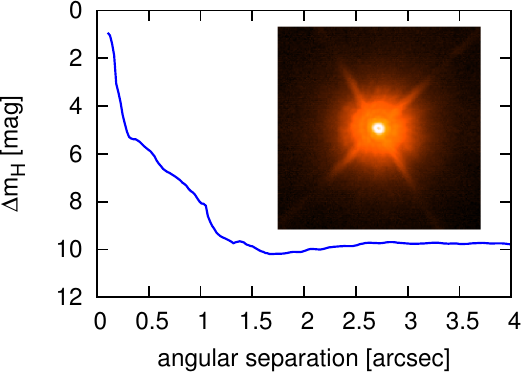}
\caption{$H$ band $5\sigma$ contrast curve for EPIC~228732031 based on the saturated image
obtained with Subaru/IRCS. The inset displays $4\arcsec\times 4\arcsec$ image
of EPIC~228732031. These data exclude companions down to a contrast of $6\times10^{-4}$
at a separation of $1\arcsec$.}
\label{ircs}
\end{figure}

We reduced the IRCS raw data as described by \citet{Hirano2016}. We applied
bias subtraction, flat-fielding, and distortion corrections before aligning and 
median-combining each of the saturated and unsaturated frames.
The FWHM of the combined unsaturated image was $0\farcs 10$.
The combined saturated image exhibits no bright source within the field of view of $21\farcs0\times 21\farcs0$. 
To estimate the achieved flux contrast, we convolved the combined saturated 
image with a kernel having a radius equal to half the FWHM. We then computed the scatter 
as a function of radial separation from EPIC~228732031. Figure \ref{ircs} 
shows the resulting $5\sigma$ contrast curve, along with a zoomed-in image of
EPIC~228732031 with a field of view of $4\farcs0\times 4\farcs0$. We can exclude companions brighter than $6\times10^{-4}$ of the target star, over separations of 1-4$\farcs$. 

\section{Stellar Parameters}

\label{sec:stellar}

We determined the spectroscopic parameters of EPIC 228732031 from the co-added HARPS-N spectrum, which has a SNR per pixel of about 165 at 5500\,\AA. We used four different methods to extract the spectroscopic parameters:

\emph{-- Method 1}. We used the spectral synthesis code \texttt{SPECTRUM}\footnote{\url{www.appstate.edu/~grayro/spectrum/spectrum.html}.} \citep[V2.76;][]{Gray1994} to compute synthetic spectra using \texttt{ATLAS\,9} model atmospheres \citep{Castelli2004}. We adopted the calibration equations of \citet{Bruntt2010} and \citet{Doyle2014} to derive the microturbulent (\vmic) and macroturbulent (\vmac) velocities. We focused on spectral features that are most sensitive to varying photospheric parameters. Briefly, we used the wings of the H$\alpha$ line to obtain an initial estimate of the effective temperature (\teff). We then used the Mg\,{\sc i}~5167, 5173, 5184~\AA, the Ca\,{\sc i}~6162, 6439~\AA, and the Na\,{\sc i}~D lines to refine the effective temperature and derive the surface gravity (\logg). The iron abundance \met\ and projected rotational velocity \vsini\ were estimated by fitting many isolated and unblended iron lines. The results were: \teff\ = $5225 \pm 70$ K; \logg\ =  $4.67 \pm 0.08$ (cgs); \met =  $0.01\pm 0.05$ dex; \vsini\ =  $4.8 \pm 0.6$ km~s$^{-1}$; \vmic =  $0.86 \pm 0.10$ km~s$^{-1}$ and \vmac =  $2.07 \pm 0.48$ km~s$^{-1}$.

\emph{-- Method 2}. We also determined the spectroscopic parameters using the equivalent-width method. The analysis was carried out with {\tt iSpec} \citep{Blanco-Cuaresma2014}. The effective temperature \teff\, surface gravity \logg\, metallicity \met\ and microturbulence \vmic\ were iteratively determined using 116 Fe I and 15 Fe II lines by requiring excitation balance, ionization balance and the agreement between Fe I and Fe II abundances. Synthetic spectra were calculated using MOOG \citep{Sneden1973} and MARCS model atmospheres \citep{Gustafsson2008}. The projected rotation velocity \vsini\ was determined by convolving the synthetic spectrum with a broadening kernel to match the observed spectrum. The results were: \teff\ = $5216 \pm 27$ K; \logg\ =  $4.63 \pm 0.05$ (cgs); \met =  $-0.02\pm 0.09$\,dex; \vsini\ =  $4.0 \pm 0.6$ km~s$^{-1}$.

\emph{-- Method 3}. We fitted the observed spectrum to theoretical \attw\ model atmospheres from \cite{Kurucz2013} using {\tt SME} version 5.22 \citep{vp96,vf05,pv2017}\footnote{\url{http://www.stsci.edu/~valenti/sme.html}.}. We used the atomic and molecular line data from VALD3 \citep{pk1995,kp1999}\footnote{\url{http://vald.astro.uu.se}.}. We used the empirical calibration equations for Sun-like stars from \cite{Bruntt2010b} and \cite{Doyle2014} to determine the microturbulent (\vmic) and macroturbulent (\vmac) velocities. The projected stellar rotational velocity \vsini\ was estimated by fitting about 100 clean and unblended metal lines. To determine the \teff\, the \halpha\ profile was fitted to the appropriate model \citep{fuhrmann93,axer94,fuhrmann94,fuhr97a,fuhr97b}. Then we iteratively fitted for \logg\ and \met\ using the Ca I lines at 6102, 6122, 6162 and 6439 \AA, as well as the Na I doublet at 5889.950 and 5895.924 \AA. The results were: \teff\ = $4975 \pm 125$ K; \logg\ =  $4.40 \pm 0.15$ (cgs); \met =  $-0.06\pm 0.10$\,dex; \vsini\ =  $4.8 \pm 1.6$ km~s$^{-1}$.

\emph{-- Method 4}.  We took a more empirical approach using {\tt SpecMatch-emp}\footnote{\url{https://github.com/samuelyeewl/specmatch-emp}.} \citep{Yee2017}. This code estimates the stellar parameters by comparing the observed spectrum with a library of about 400 well-characterized stars (M5 to F1) observed by Keck/HIRES. {\tt SpecMatch-emp} gave \teff\ = $5100 \pm 110$ K; \met =  $-0.06\pm 0.09$ and $R_\star = 0.75 \pm 0.10~R_{\odot}$. {\tt SpecMatch-emp} directly yields stellar radius rather than the surface gravity because the library stars typically have their radii calibrated using interferometry and other techniques. With the stellar radius, \teff\ and \met, we estimated the surface gravity using the empirical relation by \citet{Torres2010}: \logg =  $4.60 \pm 0.10$.

The spectroscopic parameters from these four methods do not agree with each other within the quoted uncertainties (summarized in Table \ref{spectro}), even though they are all based on the same data. In particular, the effective temperature
from Method 3 is about 2$\sigma$ lower than weighted mean of all the results.
This disagreement is typical in studies of this nature, and probably
arises because the quoted uncertainties do not include systematic effects associated with the
different assumptions and theoretical models.
For the analysis that follows, we computed the weighted mean of each spectroscopic parameter,
and assigned it an uncertainty equal to the standard deviation among the four different results. The uncertainties thus derived are likely underestimated,
because of systematic biases introduced by the various model assumptions that are difficult to quantify.
The results are:
\teff\ = $5200 \pm 100$ K; \logg\ =  $4.62 \pm 0.10$; \met =  $-0.02\pm 0.08$ and \vsini\ =  $4.4 \pm 1.0$ km~s$^{-1}$. 

We determined the stellar mass and radius using the code {\tt Isochrones} \citep{Morton2015}. This code takes as input the spectroscopic parameters, as well as the broad-band photometry of EPIC\,228732031 retrieved from the ExoFOP website\footnote{\url{https://exofop.ipac.caltech.edu}.}. The various inputs are fitted to the stellar-evolutionary models from the Dartmouth Stellar Evolution Database \citep{Dotter2008}. We used the nested sampling code {\tt MultiNest} \citep{Feroz2009} to sample the posterior distribution. The results were $M_{\star}\,=\,$0.84$\pm$0.03 $M_{\odot}$ and $R_{\star}\,=\,$0.81$\pm$0.03 $R_{\odot}$.

We derived the interstellar extinction ($A_\mathrm{v}$) and distance ($d$) to EPIC\,228732031 following the technique described in \citet{Gandolfi2008}. Briefly, we fitted the B$-$V and 2MASS colors using synthetic magnitudes extracted from the \texttt{NEXTGEN} model spectrum \citep{Hauschildt1999} with the same spectroscopic parameters as the star. Adopting the extinction law of \citet{Cardelli1989} and assuming a total-to-selective extinction of $R_\mathrm{v}=3.1$, we found that EPIC\,228732031 suffers from a small amount of reddening of $A_\mathrm{v}\,=\,0.07\pm0.05$\,mag. Assuming a black body emission at the star's effective temperature and radius,
we derived a distance from the Sun of $d\,=\,174\pm20$\,pc.

\section{Photometric Analysis}
\label{sec:photometric}

\subsection{Transit Detection}\label{transit_detection}

Before searching the {\it K2} light curve for transits, we removed long-term systematic or instrumental flux variations by fitting a cubic spline of length 1.5 days, and then dividing by the spline function.
We searched for periodic transit signals using the Box-Least-Squares algorithm \citep[BLS,][]{Kovac2002}.
Following the suggestion of \citet{Ofir2014}, we employed a nonlinear frequency grid to account for
the expected scaling of transit duration with orbital period. We also adopted his definition of signal detection efficiency (SDE), in which the significance of a detection is quantified by first subtracting the local median of the BLS spectrum and then normalizing by the local standard deviation. The transit signal of EPIC~228732031b was detected with a SDE of 14.4.

We searched for additional transiting planets in the system by re-running the BLS algorithm after removing the data within 2~hours of each transit of planet b.  No significant transit signal was detected: the maximum SDE of the new BLS spectrum was 4.5. Visual inspection of the light curve also did not reveal any significant transit events.
In particular no transit was seen at the orbital period of 3.0 days, the period which emerged as the dominant peak in the periodogram of the radial velocity data (See Section \ref{possible_planetc}). The upper panel of Figure~\ref{fig:K2_lc} shows the light curve after removing the transits of planet b.

\subsection{Transit Modeling}\label{transit_modeling}

The orbital period, mid-transit time, transit depth and transit duration from BLS were used as the starting point for a more rigorous transit analysis. We modeled the transit light curves with the {\tt Python} package {\tt Batman} \citep{Kreidberg2015}.  We isolated the transits using a 4-hour window around the time of mid-transit.  The free parameters included in the model were the orbital period $P_{\text{orb}}$, the mid-transit time $t_{\text{c}}$, the planet-to-star radius ratio $R_{\text{p}}/R_\star$; the scaled orbital distance $a/R_\star$; and the impact parameter $b\equiv a\cos i/R_\star$. We adopted a quadratic limb-darkening profile. We imposed Gaussian priors on the limb-darkening coefficients $u_1$ and $u_2$ with the median from EXOFAST\footnote{\url{astroutils.astronomy.ohio-state.edu/exofast/limbdark.shtml}.} \citep[$u_1$= 0.52, $u_2$ = 0.19, ][]{Eastman2013} and widths of 0.1. Jeffreys priors were imposed on $P_{\text{orb}}$, $R_p/R_\star$, and $a/R_\star$. Uniform priors were imposed on $t_{\text{c}}$ and $\cos\,i$. Since the data were obtained with 30~min averaging, we sampled the model light curve at 1~min intervals and then averaged to 30~min to account for the finite integration time \citep{Kipping2010}.

\begin{figure}
\begin{center}
\includegraphics[width = 1.\columnwidth]{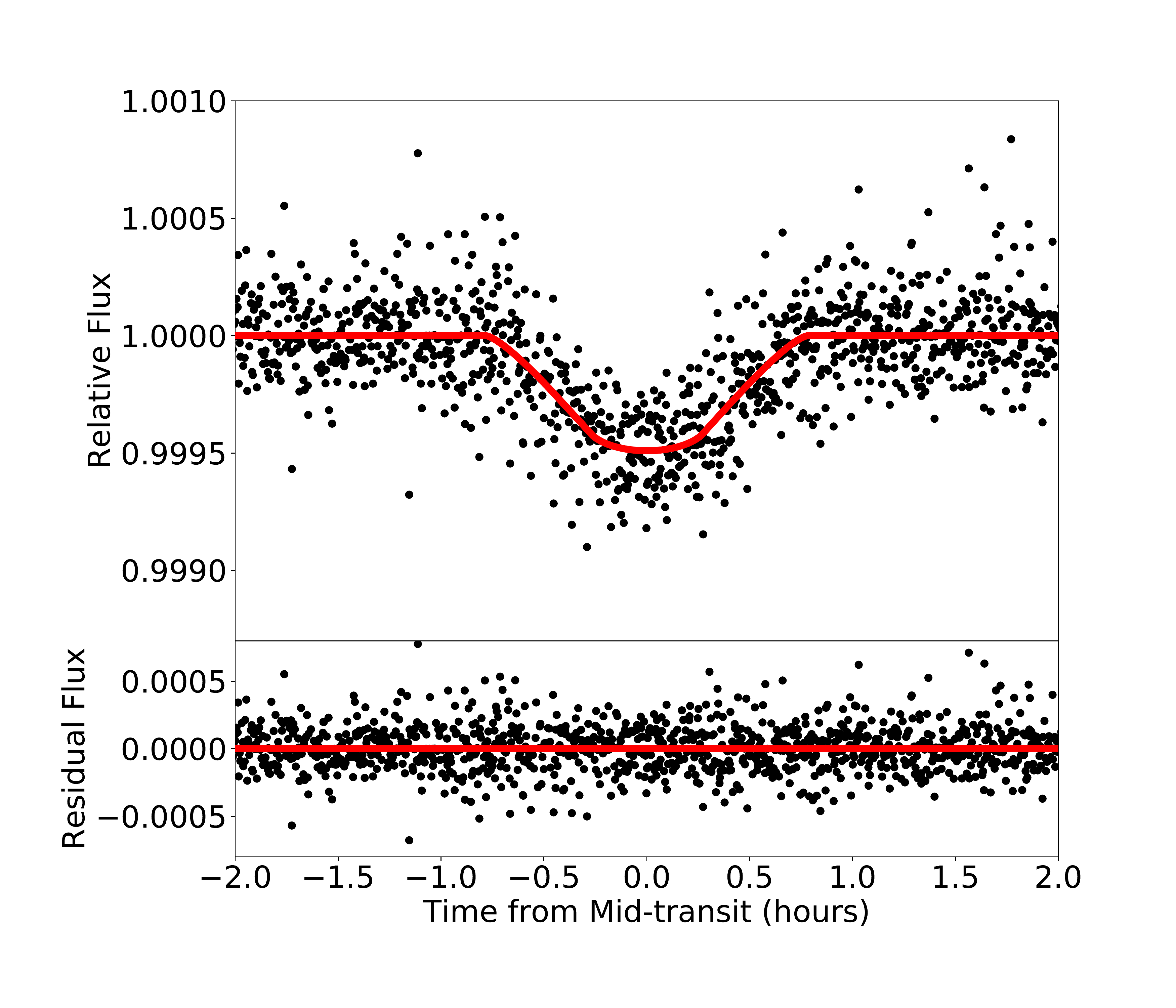}
\caption{The best-fit transit model of EPIC~228732031b. The black dots are {\it K2} observations. The red line is the best-fit transit model after accounting for the effect of the 30-min time averaging.}
\label{transit}
\end{center}
\end{figure}

\begin{figure}
\begin{center}
\includegraphics[width = 1.\columnwidth]{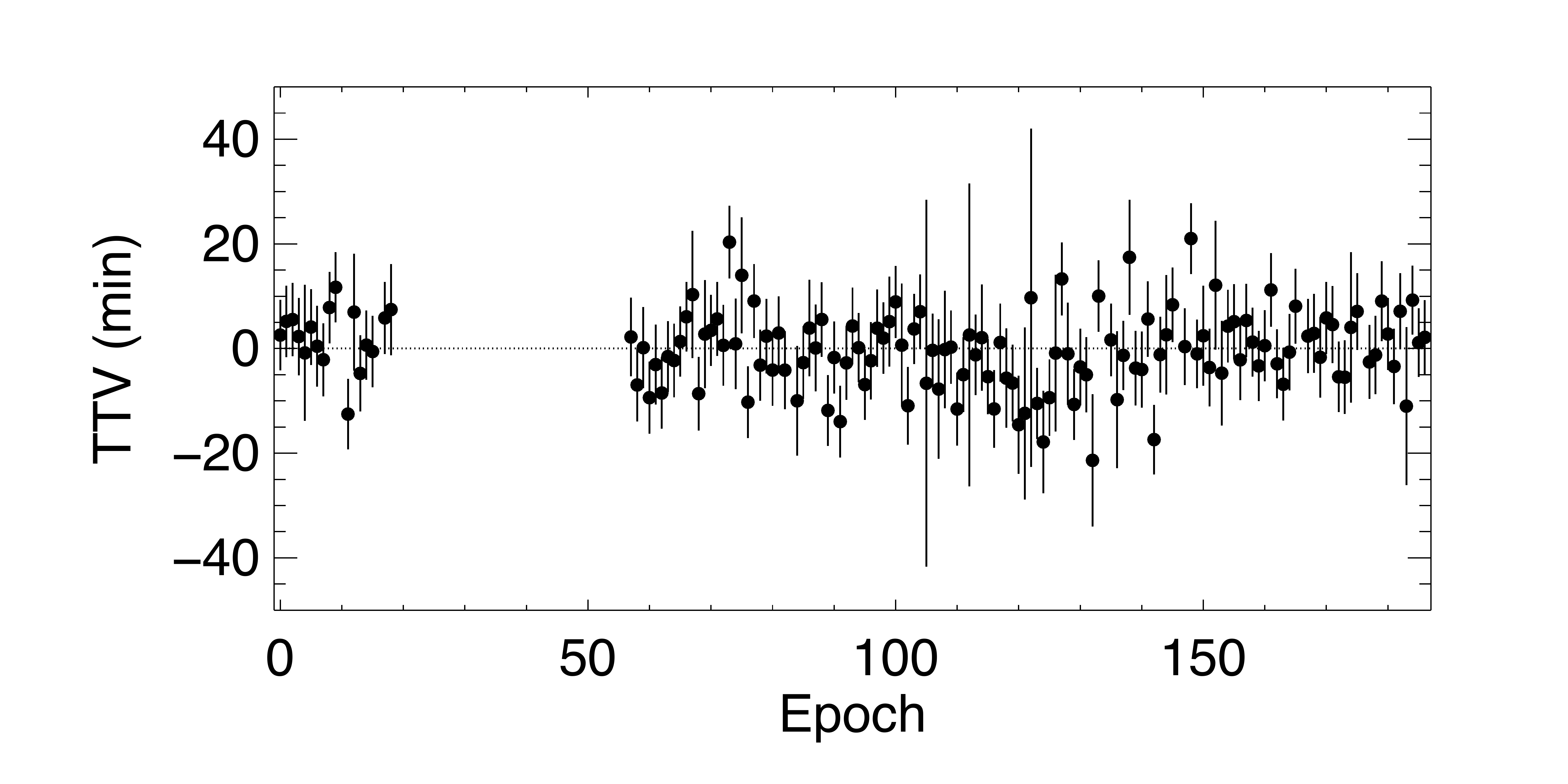}
\caption{The transit time variations of EPIC~228732031b observed by {\it K2}. They have large uncertainties due to the combination of 30-min time-averaging and the short transit duration of $\approx$1 hour. The transit times are consistent with a constant period.}
\label{ttv}
\end{center}
\end{figure}

We adopted the usual $\chi^2$ likelihood function and found the best-fit solution using the Levenberg-Marquardt algorithm implemented in the {\tt Python} package {\tt lmfit}. Figure~\ref{transit} shows the phase-folded light curve and the best-fitting model.  In order to test if planet b displays transit timing variations (TTV), we used the best-fit transit model as a template. We fitted each individual transit, varying only the mid-transit time and a quadratic function of time to describe any residual long-term flux variation. The resultant transit times are consistent with a constant period (Fig.\,\ref{ttv}).  We proceeded with the analysis under the assumption that any TTVs are negligible given the current sensitivity.

To sample the posterior distribution of various transit parameters, we performed an MCMC analysis with {\tt emcee} \citep{emcee}.  We launched 100 walkers in the vicinity of the best-fit solution. We stopped the walkers after running 5000 links and discarded the first 1000 links. Using the remaining links, the Gelman-Rubin potential scale reduction factor was found to be within 1.03, indicating adequate convergence. The posterior distributions for all parameters were smooth and unimodal. Table~\ref{planet} reports the results,
based on the 16, 50, and 84\,\% levels of the cumulative posterior distribution.
The mean stellar density obtained from transit modeling assuming a circular orbit (2.43$^{+0.61}_{-1.09}$\,g\,cm$^{-3}$) agrees with that computed from the mass and radius derived in Section~\ref{sec:stellar} (2.23$ \pm 0.33$\,g\,cm$^{-3}$).

\subsection{Stellar Rotation Period}

The {\it K2} light curve showed quasi-periodic modulations that are likely associated with magnetic activity coupled with stellar rotation (see upper panel of Fig.~\ref{fig:K2_lc}). To measure the stellar rotation period, we computed the Lomb-Scargle Periodogram \citep{Lomb1976,Scargle1982} of the {\it K2} light curve, after removing the transits of planet b. The strongest peak is at 9.37$\pm$1.85 days.  Computing the autocorrelation function \citep{McQuillan2014} leads to a consistent estimate for the stellar rotation period of $9.2^{+2.3}_{-1.2}$ days. Analysis of the ground-based AIT light curve also led to a
consistent estimate of $9.84 \pm 0.80$ days. The amplitude of the rotationally-modulated variability was about 0.5\% in both datasets.

Using the measured values of $P_{\rm rot}$, $R_\star$, and $v\sin i_\star$, it is possible to check for a large spin-orbit misalignment along the line of sight. Our spectroscopic analysis gave $v\sin i_\star = 4.4 \pm 1.0$~km~s$^{-1}$.
Using the stellar radius and rotation period reported in Table~\ref{stellar},
$v = 2\pi R_\star / P_{\rm rot} = 4.4 \pm 1.1$~km~s$^{-1}$.
Because these two values are consistent, there is no evidence for any misalignment,
and the 2$\sigma$ lower limit on $\sin i_\star$ is 0.48.

\begin{figure*}
\begin{center}
\includegraphics[width = 2.\columnwidth]{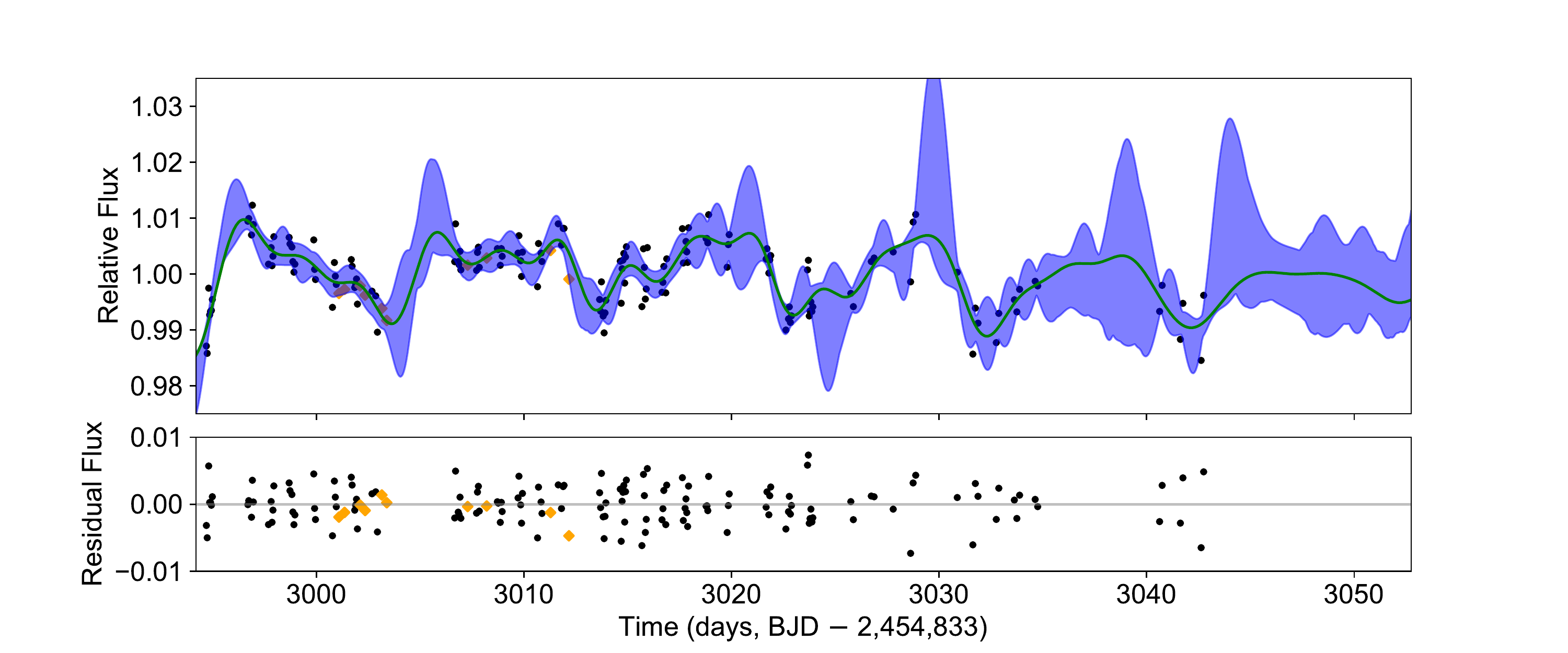}
\caption{Ground-based light curve of EPIC~228732031. The black circles are observed fluxes from AIT. The orange diamonds are observed fluxes from Swope. The light curve shows a rotational modulation with similar periodicity as the {\it K2} light curve (See Fig. \ref{ground_folded}). The green curve shows the Gaussian Process regression of the light curve with the same quasi-periodic kernel as in Fig. \ref{fig:K2_lc}. The blue shaded region is the 1$\sigma$ confidence interval of the Gaussian Process.}
\label{AIT}
\end{center}
\end{figure*}

\begin{figure}
\begin{center}
\includegraphics[width = 1.\columnwidth]{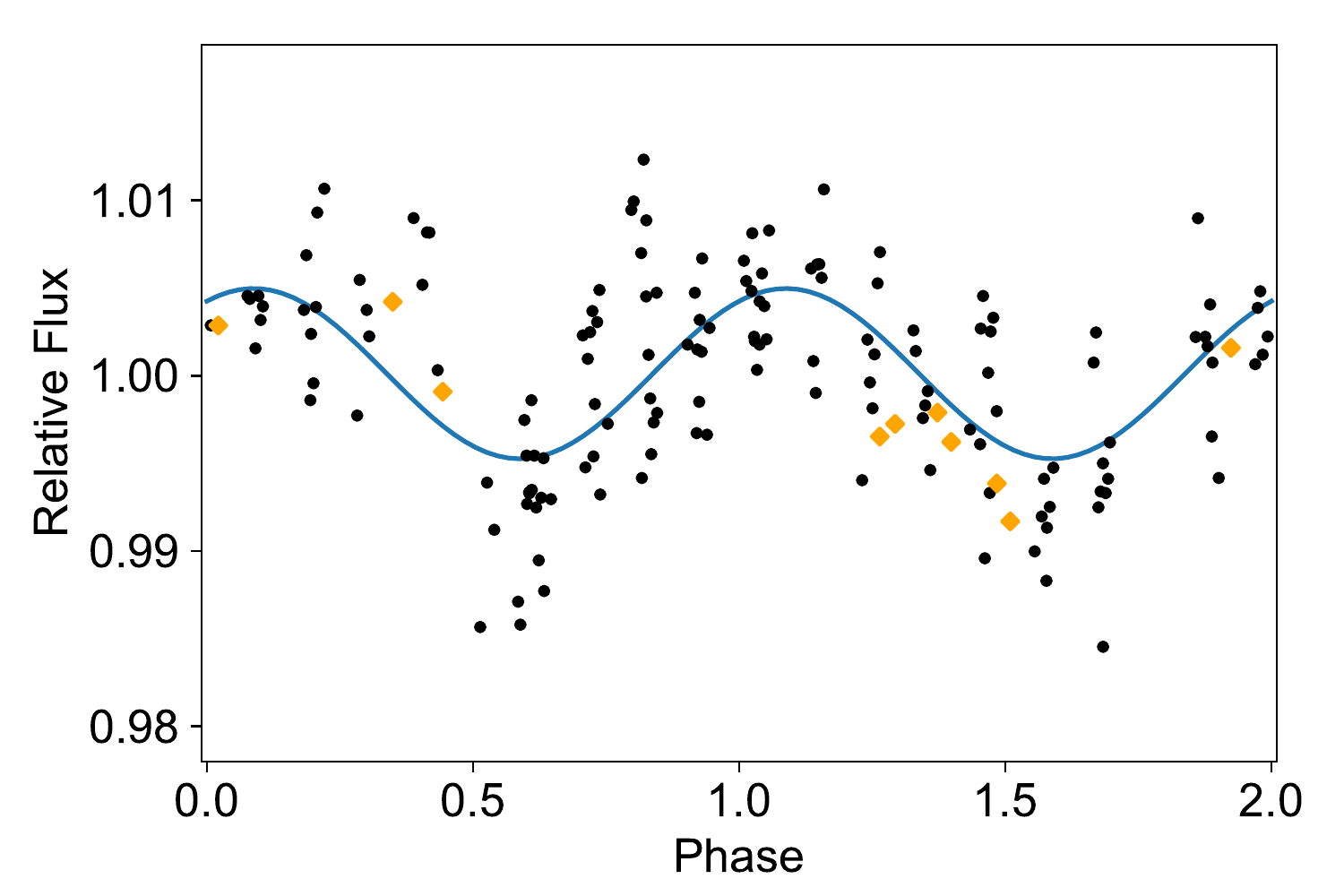}
\caption{Ground-based light curves of EPIC~228732031, as a function of stellar rotational phase. The black circles are from AIT. The orange diamonds are from Swope. The blue line is a sinusoidal fit.}
\label{ground_folded}
\end{center}
\end{figure}

\section{Radial Velocity Analysis}\label{rv_analysis}

Stellar variability is a frequent source of correlated noise in precise RV data. Stellar variability may refer to several effects including $p$-mode oscillations, granulation, magnetic activity coupled with stellar rotation, and long-term magnetic activity cycles. The most problematic component is often the magnetic activity coupled with stellar rotation. The magnetic activity of a star gives rise to surface inhomogeneities: spots, plages, and faculae. As these active regions are carried around by the rotation of the host star, they produce two major effects on the radial velocity measurement \citep[see, e.g.,][]{Lindegren2003,Haywood2016}. (1) The "Rotational" component: stellar rotation carries the surface inhomogeneities from the blue-shifted to the red-shifted part of the star, distorting the spectral lines and throwing off the
apparent radial velocity. (2) The "Convective" component: the suppression of convective blueshift in strong magnetic regions leads to a net radial velocity shift whose amplitude depends on the orientation of the surface relative to the observer's line of sight. Both of these effects produce quasi-periodic variations in the radial velocity measurements on the timescale of the stellar rotation period.

\begin{figure*}
\begin{center}
\includegraphics[width = 0.5\columnwidth]{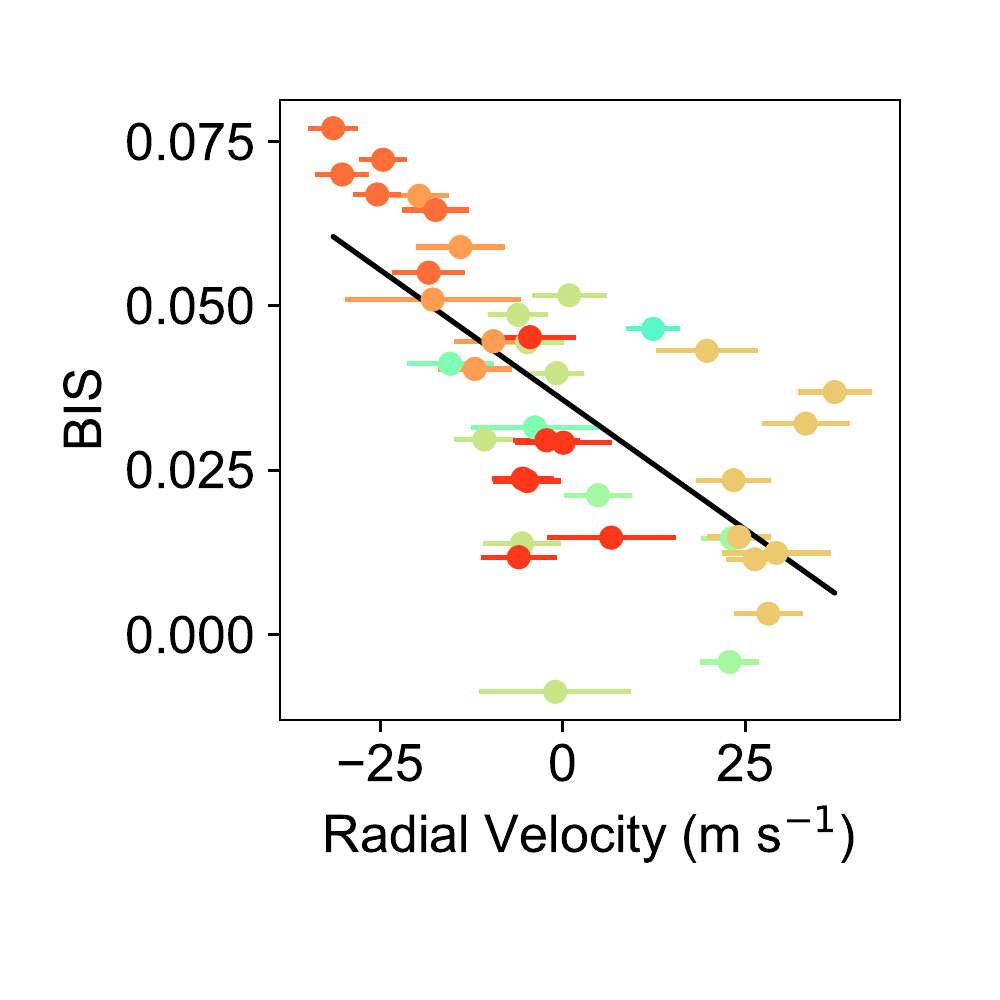}
\includegraphics[width = .5\columnwidth]{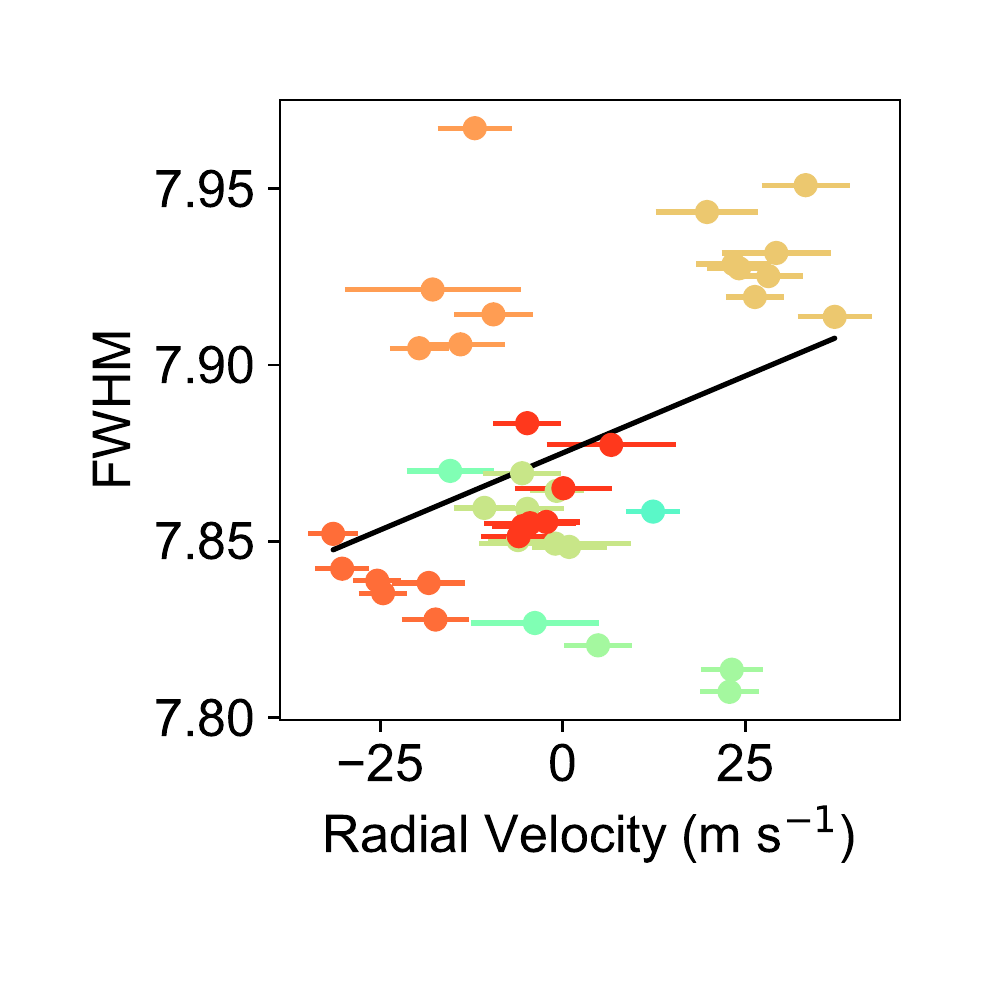}
\includegraphics[width = .5\columnwidth]{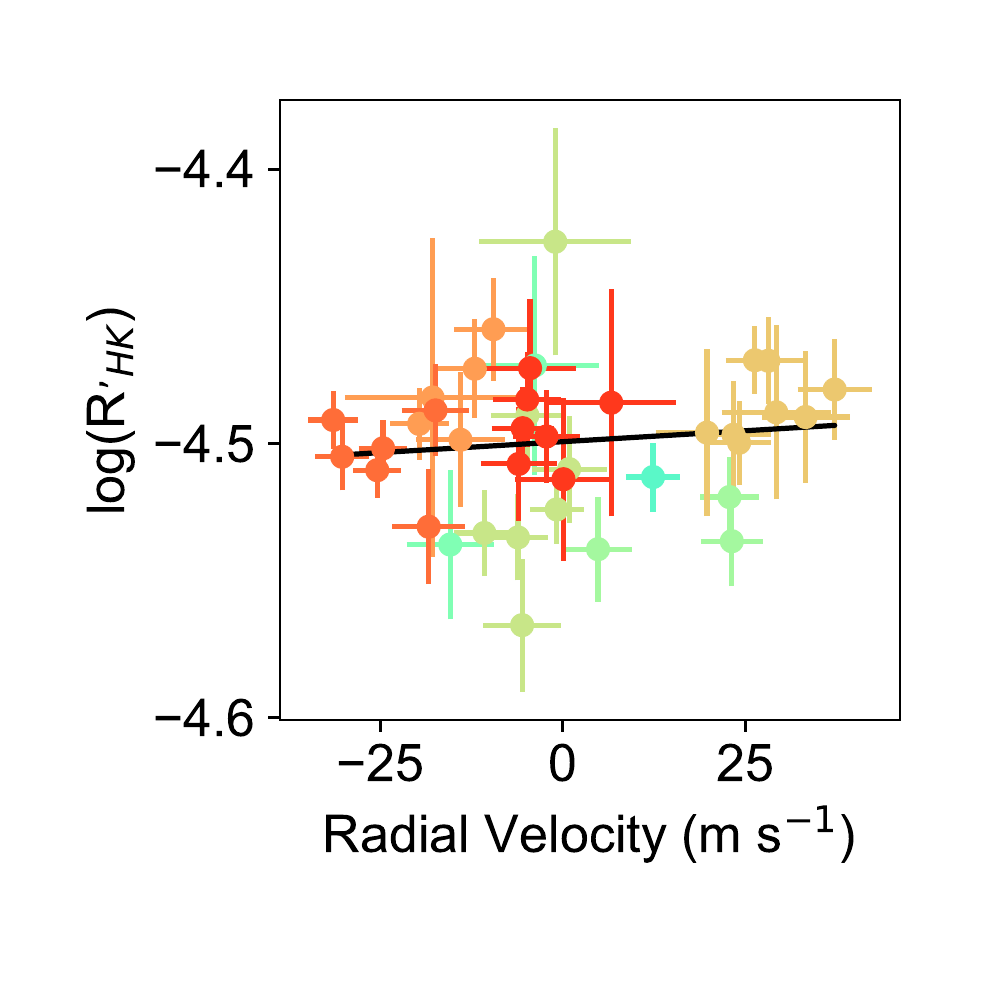}
\includegraphics[width = .5\columnwidth]{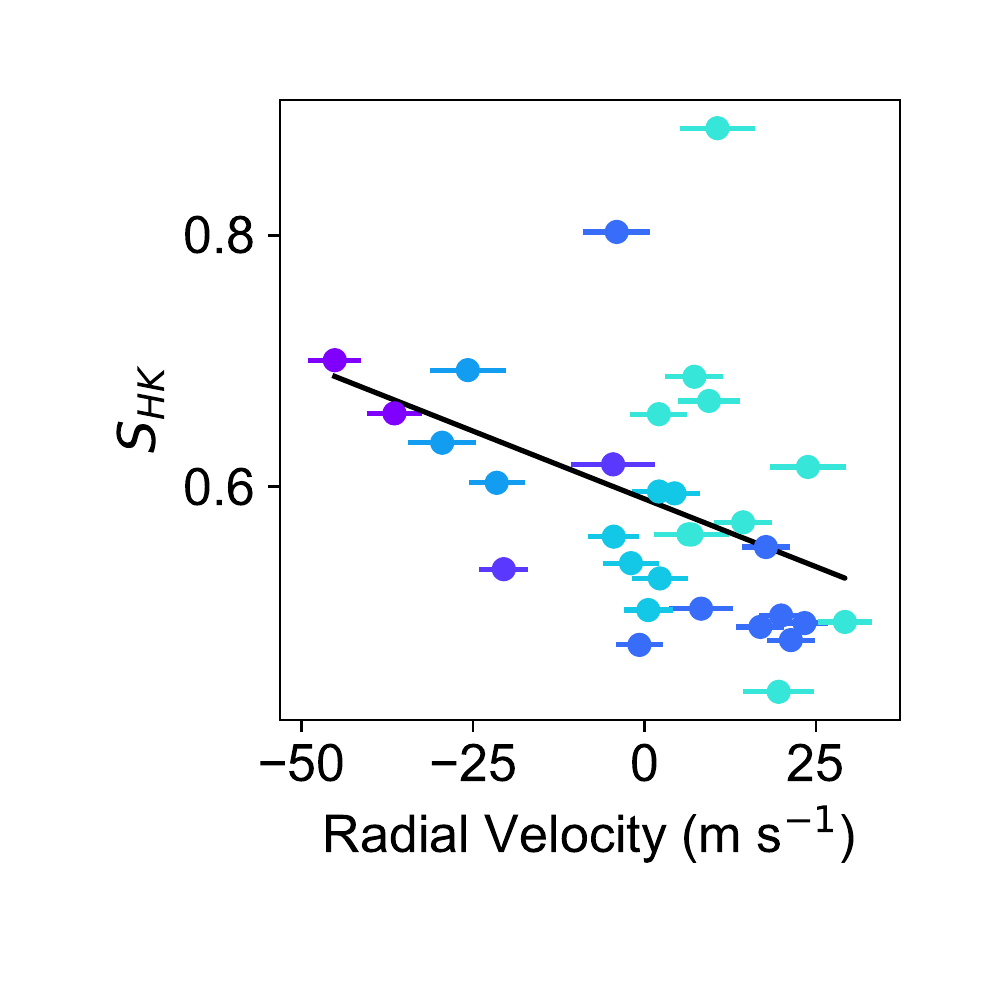}
\includegraphics[width = 0.5\columnwidth]{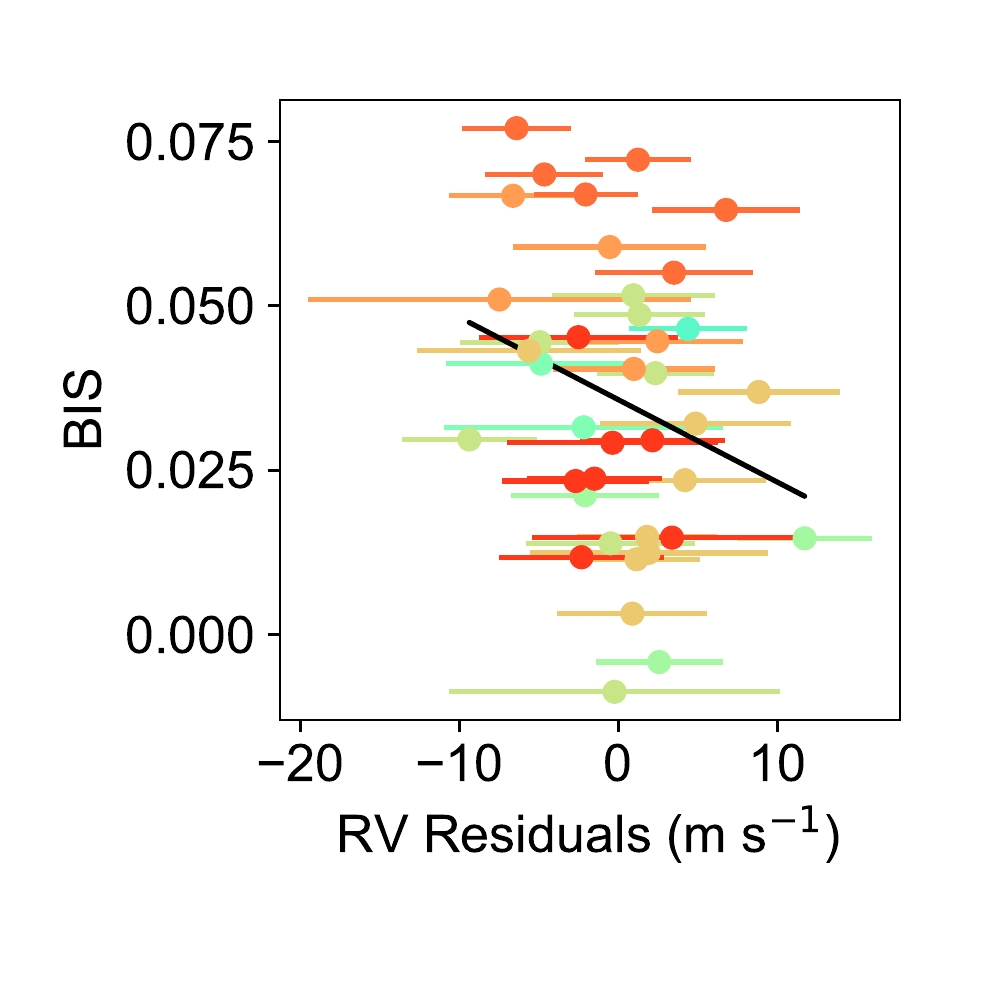}
\includegraphics[width = .5\columnwidth]{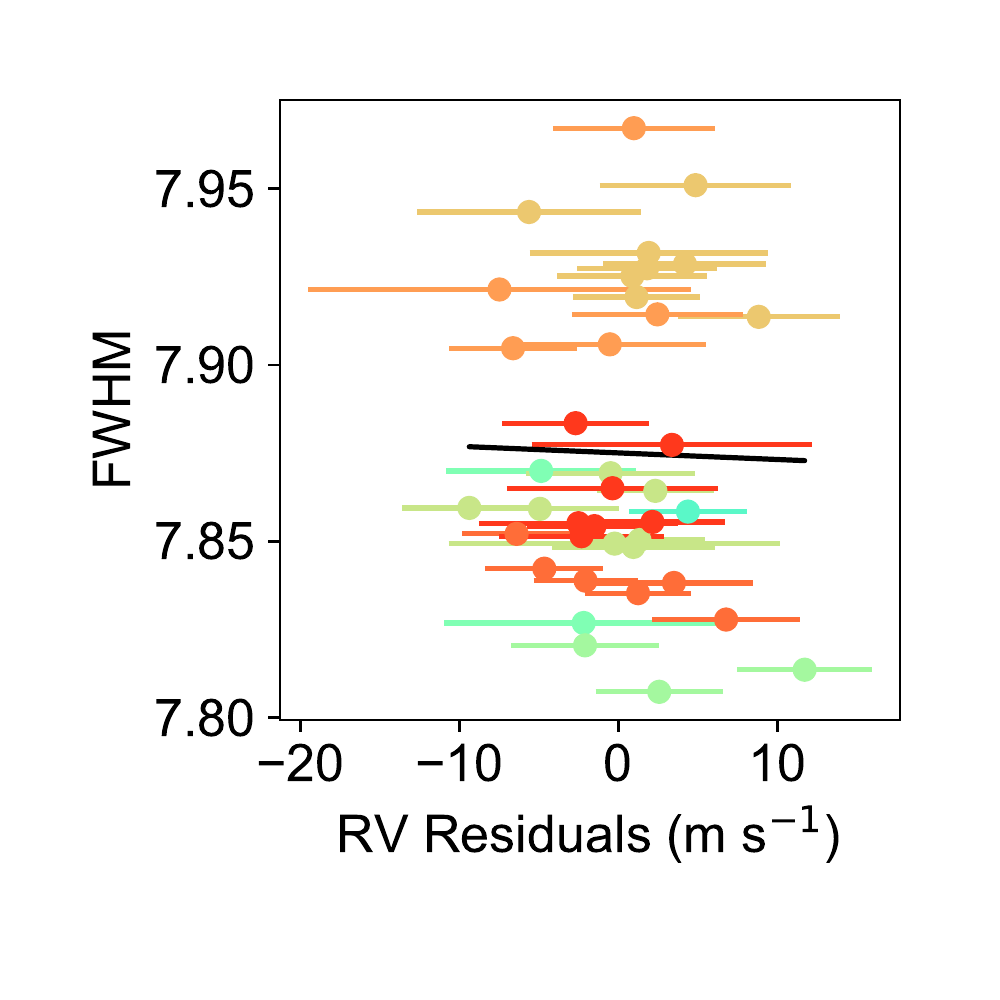}
\includegraphics[width = .5\columnwidth]{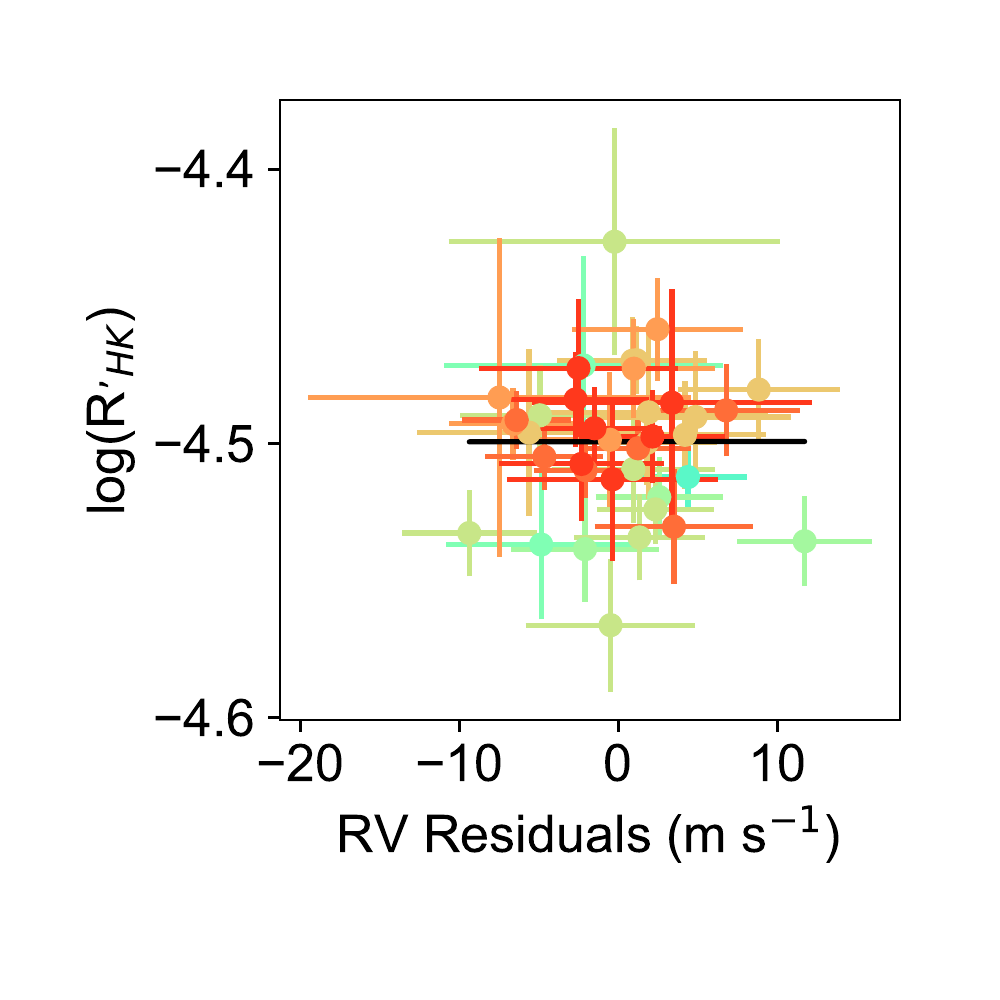}
\includegraphics[width = .5\columnwidth]{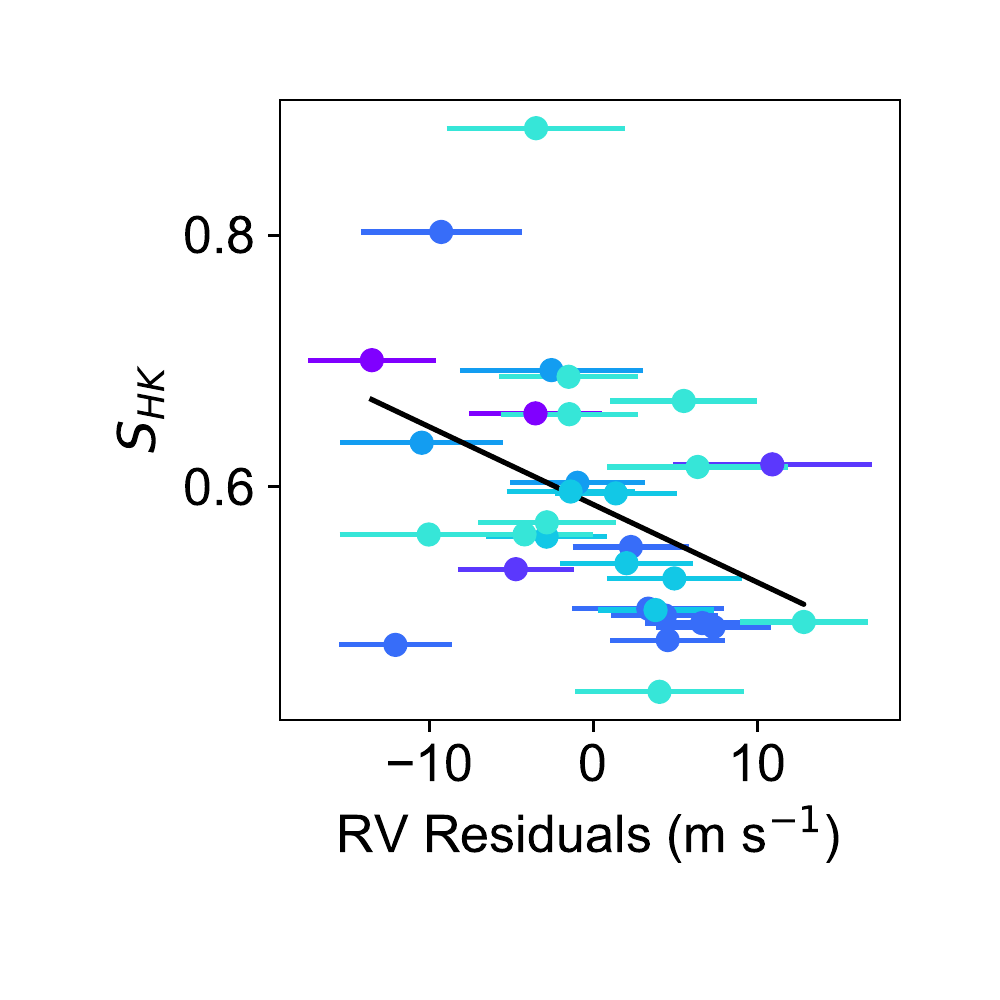}

\caption{{\it Top.}---Measured RV plotted against various stellar activity indicators. Observations from different nights are plotted in different colors. The black line is the best-fit linear correlation between RV and the activity indicator. Among these, BIS, FWHM and $S_{\text{HK}}$ showed statistically significant correlation with the measured RV with Pearson $p$-values of 2.4$\times 10^{-6}$, 0.014 and 0.027 respectively. {\it Bottom.}---Same, but using the residual RV
after removing the Gaussian Process regression model. This model largely succeeded in removing the correlations between RV and activity indicators. None of the activity indicators showed statistically significant correlation with the measured RV.}
\label{correlations}
\end{center}
\end{figure*}

The median value of log\,$R^\prime_\mathrm{HK}$ for EPIC~228732031 was $-4.50$. This suggests a relatively strong chromospheric activity level, according to \citet{Isaacson2010}. For comparison, \citet{Egeland2017} measured a mean $\log R^\prime_\mathrm{HK}$ of about $-4.96$ for the Sun during solar cycle 24. According to \citet{Fossati2017}, the measured log\,$R^\prime_\mathrm{HK}$ is likely suppressed by the Ca II lines in the interstellar medium. 
The star might be more active than what the measured log\,$R^\prime_\mathrm{HK}$ suggests.
The amplitude of the rotational modulation seen in the photometry is well in excess of the Sun's variability.
Figure~\ref{correlations} shows the measured RV, plotted against activity
indicators. The different colors represents data obtained on different
nights. The data from different nights tend to cluster
together in these plots. This implies that the pattern of stellar activity changes on a nightly basis, and that the RVs are correlated with stellar activity. To quantify the significance of the correlations, we applied the Pearson correlation test to each activity indicator. BIS, FWHM and $S_{\text{HK}}$ showed the strongest correlations with $p$-values of 2.4$\times 10^{-6}$, 0.014 and 0.027 respectively. Both the PFS and HARPS-N data were affected by correlated noise. In order to extract the planetary signal we used two different approaches: the Floating Chunk Offset Method and Gaussian Process Regression, as described below.

\subsection{Floating Chunk Offset Method}

The Floating Chunk Offset Method \citep[see, e.g.,][]{Hatzes2011} takes advantage of the clear separation of timescales between the orbital period (0.37 days) and the stellar rotation period (9.4 days). Only the changes in velocity observed within a given night are used to determine the spectroscopic
orbit, and thereby the planet mass. In practice this is done by
fitting all of the data but allowing the data from each night to have
an arbitrary RV offset. This method requires multiple observations taken within the same night, such as those presented in this paper. 

The PFS and HARPS-N data span 14 nights, thereby introducing 14 parameters: $\gamma_1$ to $\gamma_{14}$.
We fitted a model in which the orbit was required to be circular,
and another model in which the orbit was allowed to be eccentric.
The circular model has three parameters: the RV semi-amplitude $K$, the orbital period $P_{\text{orb}}$ and time of conjunction $t_{\text{c}}$. The eccentric model has two additional parameters: the eccentricity $e$ and the argument of periastron $\omega$; for the fitting process we used
$\sqrt{e}$cos$\omega$ and $\sqrt{e}$sin$\omega$.
We also included a separate "jitter" parameters $\sigma_{\text{jit}}$ for PFS and HARPS-N. The jitter parameter accounts for both time-uncorrelated astrophysical RV noise as well as instrumental noise in excess
of the internally estimated uncertainty. We imposed Gaussian priors on the orbital period and time of conjunction, based on the photometric results from Section \ref{transit_modeling}. We imposed Jeffreys priors on $K$ and $\sigma_{\text{jit}}$, and uniform priors on $\sqrt{e}$cos$\omega$ (with range [-1,1]), $\sqrt{e}$sin$\omega$ ([-1,1]) and $\gamma_1$ to $\gamma_{14}$.

We adopted the following likelihood function:
\begin{equation}\label{rv_likelihood}
\mathcal{L}=  \prod_{i}\left({\frac{1}{\sqrt{2 \pi (\sigma_i^2 + \sigma_{\text{jit}}(t_i)^2)}}}  \exp \left[ - \frac{[RV(t_i) - \mathcal{M}(t_i)-\gamma(t_i)]^2}{2 (\sigma_i^2 + \sigma_{\text{jit}}(t_i)^2)} \right] \right),
\end{equation}
where $RV(t_i)$ is the measured radial velocity at time $t_i$; $\mathcal{M}(t_i)$ is the calculated radial velocity variation at time $t_i$; $\sigma_{i}$ is the internal measurement uncertainty; $\sigma_{\text{jit}}(t_i)$ is the jitter parameter specific to the instrument used;
and $\gamma(t_i)$ is the arbitrary RV offset specific to each night.

\begin{figure}
\begin{center}
\includegraphics[width = 1.\columnwidth]{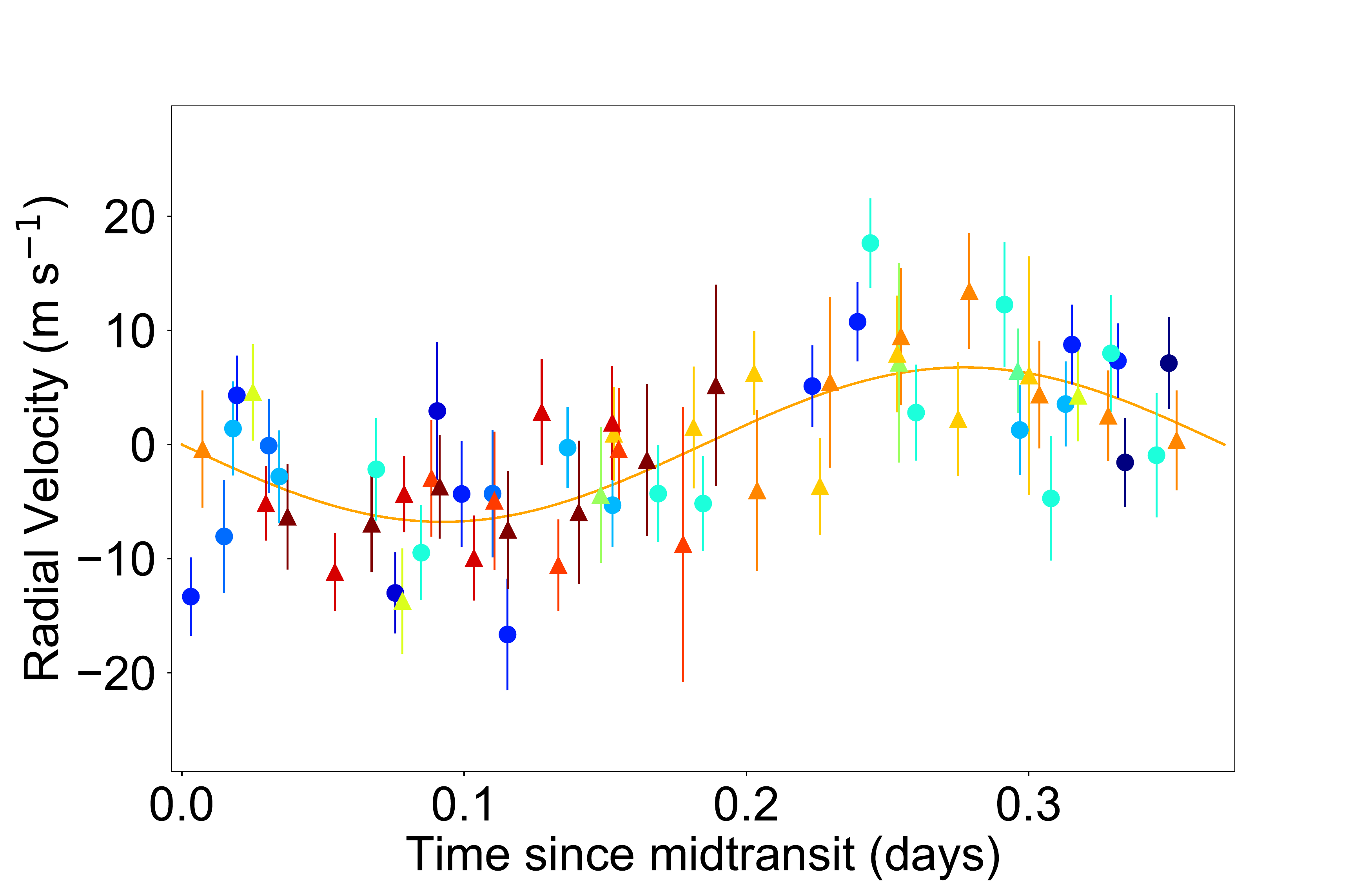}
\caption{The best-fit model assuming a circular orbit, using the Floating Chunk Method. Each color
shows the data from a single night. The circles are PFS data and the triangles are HARPS-N data. The orange line is the best-fitting model.}
\label{fco}
\end{center}
\end{figure}

We obtained the best-fit solution using the Levenberg-Marquardt algorithm implemented in the {\tt Python} package {\tt lmfit} (see Figure~\ref{fco}). To sample the posterior distribution, we performed an MCMC analysis with {\tt emcee} following a similar procedure as described in Section \ref{transit_modeling}.
Table~\ref{planet} gives the results. In the circular model, the RV semi-amplitude of planet b is $6.77 \pm 1.50$\,m s$^{-1}$ which translates into a planetary mass of 6.8\,$\pm$\,1.6\,$M_{\oplus}$. The mean density of the planet is 6.3\,$^{+3.1}_{-2.8}$\,g\,cm$^{-3}$. 

In the eccentric model, $K= 6.64 \pm 1.55$m s$^{-1}$ and the eccentricity is consistent with zero,
$e < 0.26$ (95\% confidence level). We compared the circular and eccentric models using the Bayesian Information Criterion, $BIC = -2\times \text{log}(\mathcal{L}_{\text{max}}) + N~\text{log}(M)$, where $\mathcal{L}_{\text{max}}$ is the maximum likelihood, $N$ is the number of parameters and $M$ is the number of data points \citep{Schwarz1978, Liddle2007}. The circular model is favored by a $\Delta BIC=46.3$. For this reason, and because tidal dissipation is expected to circularize such a short-period orbit,
in what follows we adopt the results from the circular model.

\subsection{Gaussian Process}
\label{sec:gp}

A Gaussian Process is a model for a stochastic process in which a parametric form is
adopted for the covariance matrix.  Gaussian Processes have been used to model the correlated noise in the RV datasets for several exoplanetary systems \citep[e.g. ][]{Haywood2014,Grunblatt2015,LM2016}. Following \citet{Haywood2014}, we chose a quasiperiodic kernel:
\begin{equation}
\label{covar}
C_{i,j} = h^2 \exp{\left[-\frac{(t_i-t_j)^2}{2\tau^2}-\Gamma \sin^2{\frac{\pi(t_i-t_j)}{T}}\right]}+\left[\sigma_i^2+\sigma_{\text{jit}}(t_i)^2\right]\delta_{i,j}
\end{equation}
where $C_{i,j}$ is an element of the covariance matrix, $\delta_{i,j}$ is the Kronecker delta function,
$h$ is the covariance amplitude, $t_i$ is the time of $i$th observation, $\tau$ is the correlation timescale, $\Gamma$ quantifies the relative importance between the squared exponential and periodic parts of the kernel, and $T$ is the period of the covariance. $h$, $\tau$, $\Gamma$ and $T$ are the "hyperparameters" of the kernel.
We chose this form for the kernel because the hyperparameters have simple physical interpretations
in terms of stellar activity: $\tau$ and $\Gamma$ quantify the typical lifetime of active regions and $T$ is closely related to the stellar rotation period.
We also introduced a jitter term $\sigma_{\text{jit}}$ specific to each instrument, to account for astrophysical and instrumental white noise.

The corresponding likelihood function has the following form:
\begin{equation}
\label{likelihood}
\log{\mathcal{L}} =  -\frac{N}{2}\log{2\pi}-\frac{1}{2}\log{|\bf{C}|}-\frac{1}{2}\bf{r}^{\text{T}}\bf{C} ^{-\text{1}} \bf{r}
\end{equation}
where $\mathcal{L}$ is the likelihood, $N$ is the number of data points, $\bf{C}$ is the covariance matrix, and $\bf{r}$ is the residual vector (the observed RV minus the calculated value). The model includes the RV variation induced by the planet and a constant offset for each observatory.
Based on the preceding results we assumed the orbit to be circular.
To summarize, the list of parameters are: the jitter parameter and offset for each of the two spectrographs, the hyperparameters $h$, $\tau$, $\Gamma$ and $T$; and for each planet considered, its RV semi-amplitude $K$, the orbital period $P_{\text{orb}}$ and the time of conjunction $t_{\text{c}}$. If non-zero eccentricity is allowed, two more parameters were added for each planet: $\sqrt{e}$cos$\omega$ and $\sqrt{e}$sin$\omega$. Again we imposed Gaussian priors on $P_{\text{orb}}$ and $t_{\text{c}}$ for the planet b based on the fit to the transit light curve. We imposed Jeffreys priors on $h$, $K$, and the
jitter parameters. We imposed uniform priors on $\gamma_{\text{HARPS-N}}$, $\gamma_{\text{PFS}}$, $\sqrt{e}$cos$\omega$ ([-1,1]) and $\sqrt{e}$sin$\omega$ ([-1,1]). The hyperparameters $\tau$, $\Gamma$ and $T$ were constrained through Gaussian Process regression of the observed light curve, as described below.

\subsubsection{Photometric Constraints on the hyperparameters}

The star's active regions produce apparent variations in both the RV and flux. Since the activity-induced flux variation and the radial velocity variation share the same physical origin, it is reasonable that they can be described by similar Gaussian Processes \citep{Aigrain2012}. We used the {\it K2} and the ground-based photometry to constrain the hyperparameters, since the photometry has higher precision and better time sampling than the RV data.

When modeling the photometric data we used the same form for the covariance matrix (Eqn.~\ref{covar}) and the likelihood function (Eqn.~\ref{likelihood}). However, we replaced $h$ and $\sigma_{\text{jit}}$ with $h_{\text{phot}}$ and $\sigma_{\text{phot}}$ since the RV and photometric data have different units.
The residual vector $\bf{r}$ in this case designates the measured flux minus a constant flux $f_0$.
We also imposed a Gaussian prior on $T$ of $9.37 \pm 1.85$ days. We imposed Jeffreys priors on
$h_{\text{phot, K2}}$, $h_{\text{phot, AIT}}$,  $h_{\text{phot, Swope}}$, $\sigma_{\text{phot, K2}}$, $\sigma_{\text{phot, AIT}}$, $\sigma_{\text{phot, Swope}}$, $\tau$ and  $\Gamma$. We imposed uniform priors on $f_{\text{0, K2}}$, $f_{\text{0, AIT}}$ and $f_{\text{0, Swope}}$.

We found the best-fit solution using the Nelder-Mead algorithm implemented in the {\tt Python} package {\tt scipy}. Figures~\ref{fig:K2_lc} and \ref{AIT} show the best-fitting Gaussian Process regression and its
uncertainty range. To sample the posterior distributions, we used {\tt emcee}, as described in Section \ref{transit_modeling}. The posterior distributions are smooth and unimodal, leading to the
following results for the hyperparameters: $\tau$ = $9.5 \pm 1.0$ days,  $\Gamma$ = $3.32 \pm 0.58$, $T$ = $9.64 \pm 0.12$ days.  These were used as priors in the Gaussian Process analysis of the RV data.

\subsection{Mass of planet b}
\begin{figure}
\begin{center}
\includegraphics[width = 1.\columnwidth]{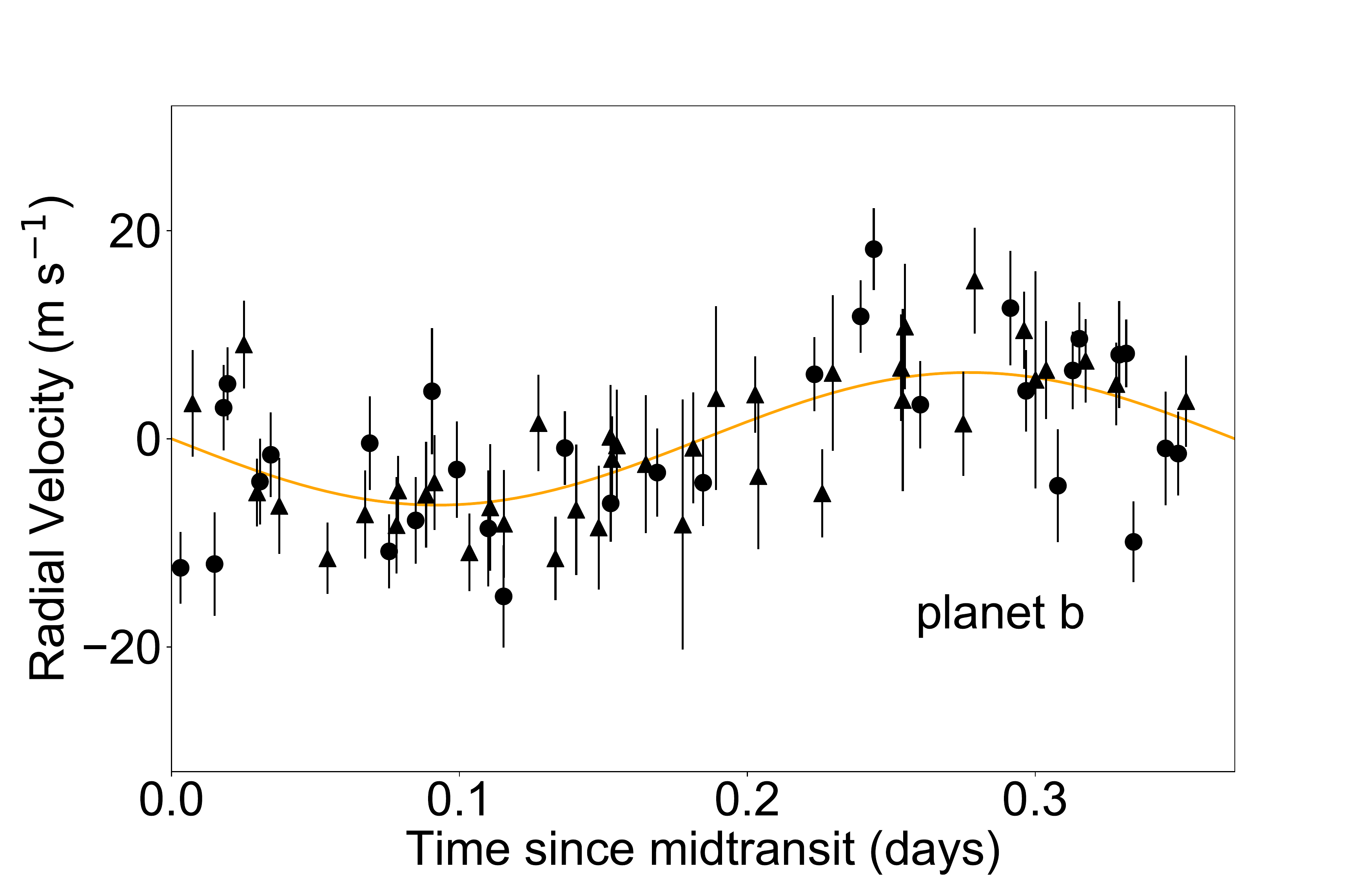}
\caption{The best-fit circular-orbit model for planet b, using Gaussian Process regression. The model for the correlated stellar noise (the blue dotted line in Figure~\ref{gp}) has been subtracted. The circles are PFS data and the triangles are HARPS-N data.}
\label{gp_fold}
\end{center}
\end{figure}

With the constraints on the hyperparameters obtained from the previous section, we analyzed the measured RV with Gaussian Process regression. We found the best-fit solution using the Nelder-Mead algorithm implemented in the {\tt Python} package {\tt scipy}. Allowing for a non-zero eccentricity did not lead to an improvement in the BIC,
so we assumed the orbit to be circular for the subsequent analysis. We sampled the parameter posterior distribution, again using {\tt emcee}, giving smooth and unimodal distributions. Table~\ref{planet} reports
the results, based on the 16, 50, and 84\% levels of the distributions. The RV semi-amplitude for planet b is $6.55 \pm 1.48$~m~s$^{-1}$, which is consistent with that obtained with the Floating Chunk Method.
This translates into a planetary mass of $6.5\pm1.6$~$M_{\oplus}$
and a mean density of $6.0^{+3.0}_{-2.7}$~g~cm$^{-3}$. Fig. \ref{gp_fold} shows the signal of planet b after
removing the correlated stellar noise.

\begin{figure*}
\begin{center}
\includegraphics[width = 1.5\columnwidth]{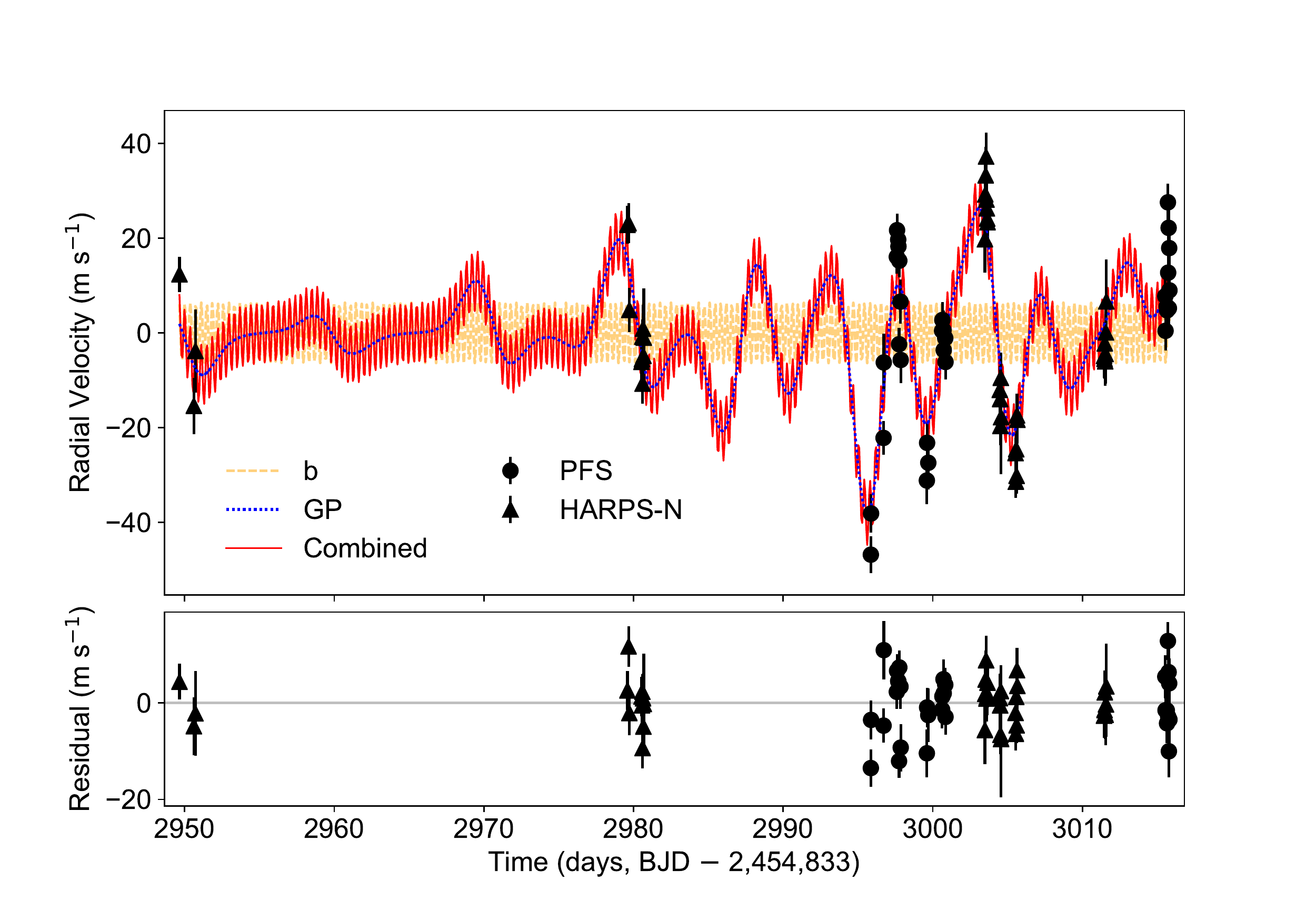}
\caption{The measured radial velocity variation of EPIC~228732031 and the best-fit Gaussian Process
model. The circles are PFS data and the triangles are HARPS-N data. The red solid line is the best-fit model including the signal of planet b and the correlated stellar noise. The yellow dashed line is the
signal of planet b. The blue dotted line is the Gaussian Process.}
\label{gp}
\end{center}
\end{figure*}

Fig. \ref{gp} shows the measured RV variation of EPIC~228732031 and the Gaussian Process regression. The correlated noise component dominates the model for the
observed radial velocity variation. The amplitude of the correlated noise is $h_{\text{rv}}$ = $26.0^{+7.3}_{-5.1}$\,m\,s$^{-1}$. This is consistent with the level of correlated noise we expected from stellar
activity. Based on the observed amplitude of photometric modulation and the projected stellar rotational velocity, we expected
\begin{equation}
h_{\text{rv}} \approx v\sin i_\star \times h_{\text{phot}} = 4.4~\text{km s}^{-1} \times 0.005 \approx 22~\text{m s}^{-1}.
\end{equation}
The Gaussian Process regression successfully removed most of the correlated noise, as well as the correlations
between the measured RV and the activity indicators. This is shown in the lower panel of Figure~\ref{correlations}. The clustering of
nightly observations seen in the original dataset (upper panel)
was significantly reduced. Pearson correlation tests showed
that none of the activity indicators correlate significantly with
the radial velocity residuals.

\subsubsection{Planet c?}\label{possible_planetc}
\begin{figure*}
\begin{center}
\includegraphics[width = 1.\columnwidth]{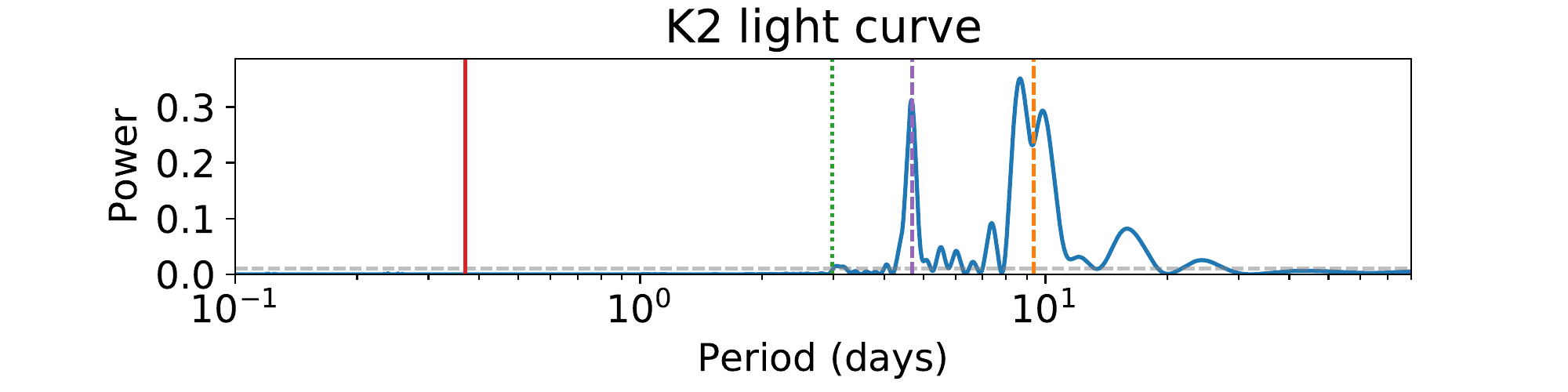}
\includegraphics[width = 1.\columnwidth]{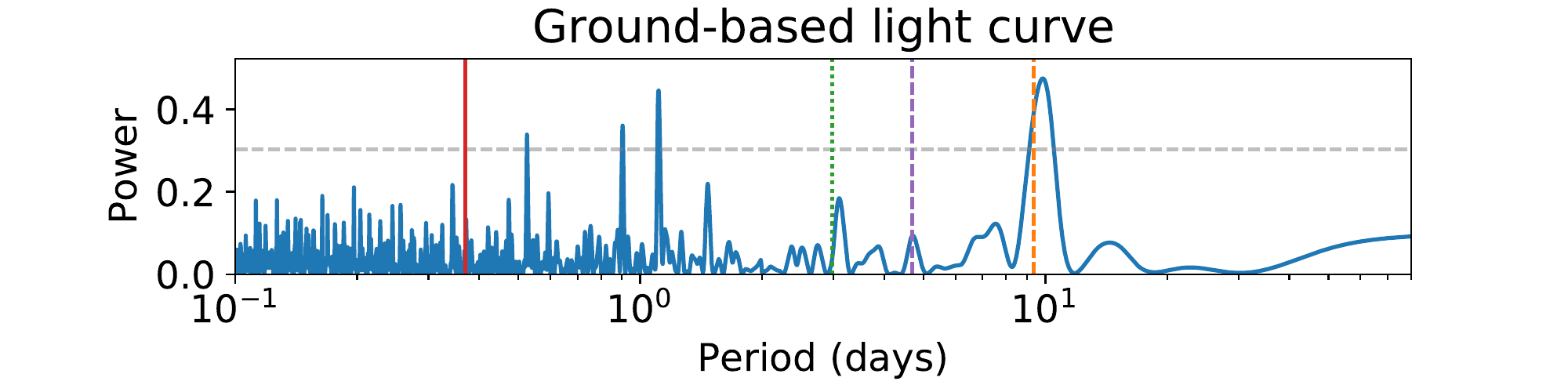}
\includegraphics[width = 1.\columnwidth]{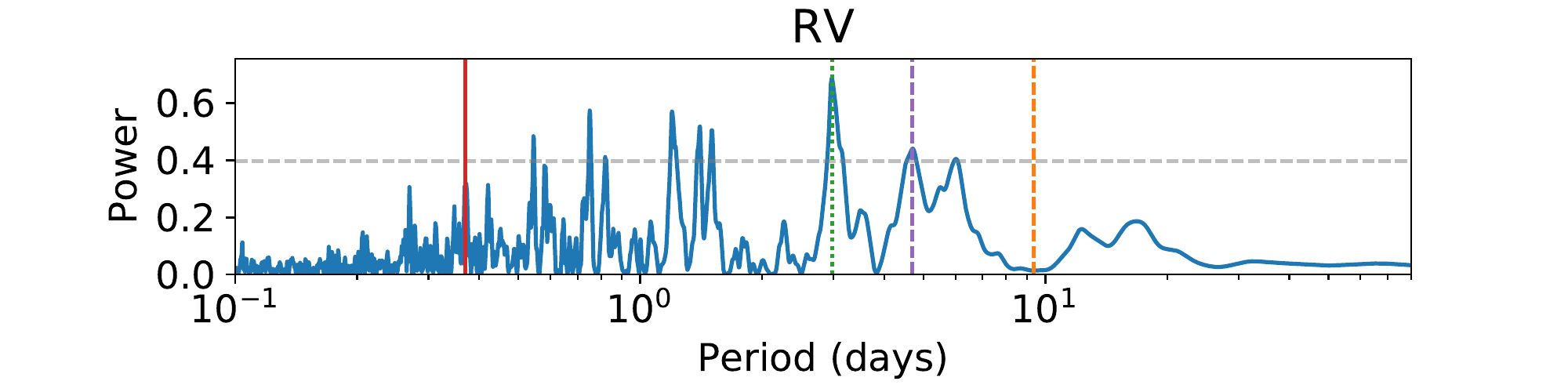}
\includegraphics[width = 1.\columnwidth]{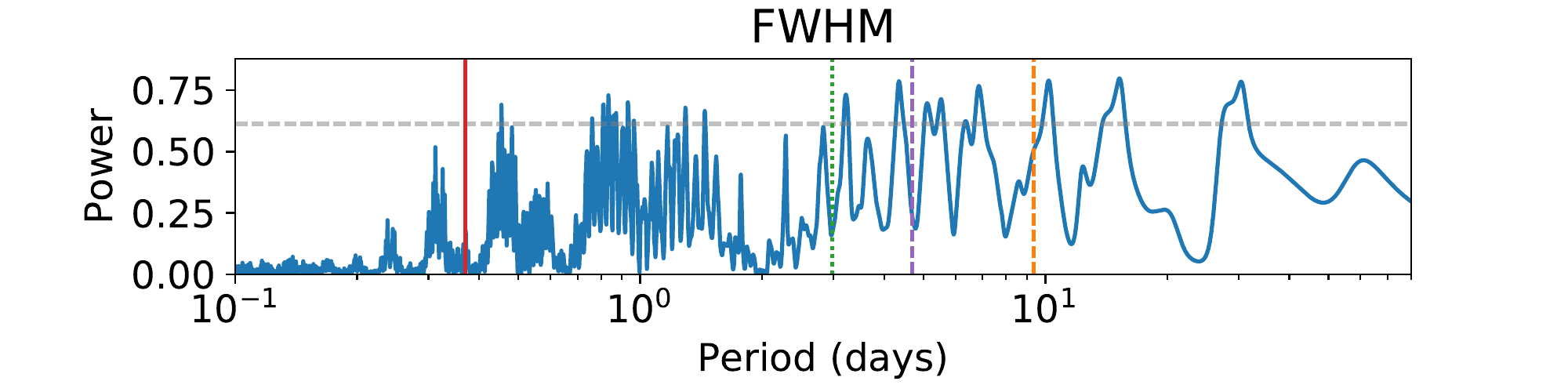}
\includegraphics[width = 1.\columnwidth]{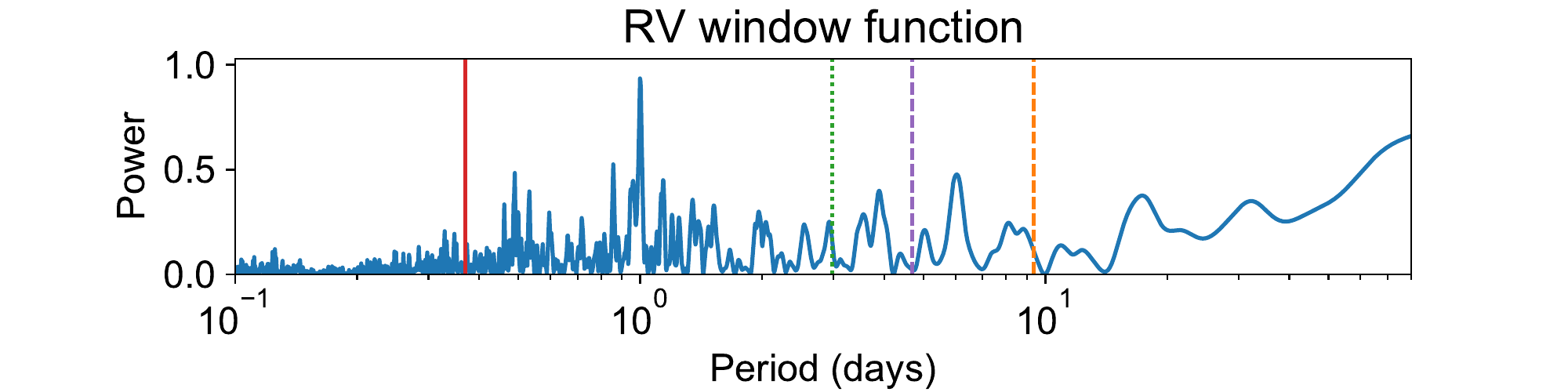}
\includegraphics[width = 1.\columnwidth]{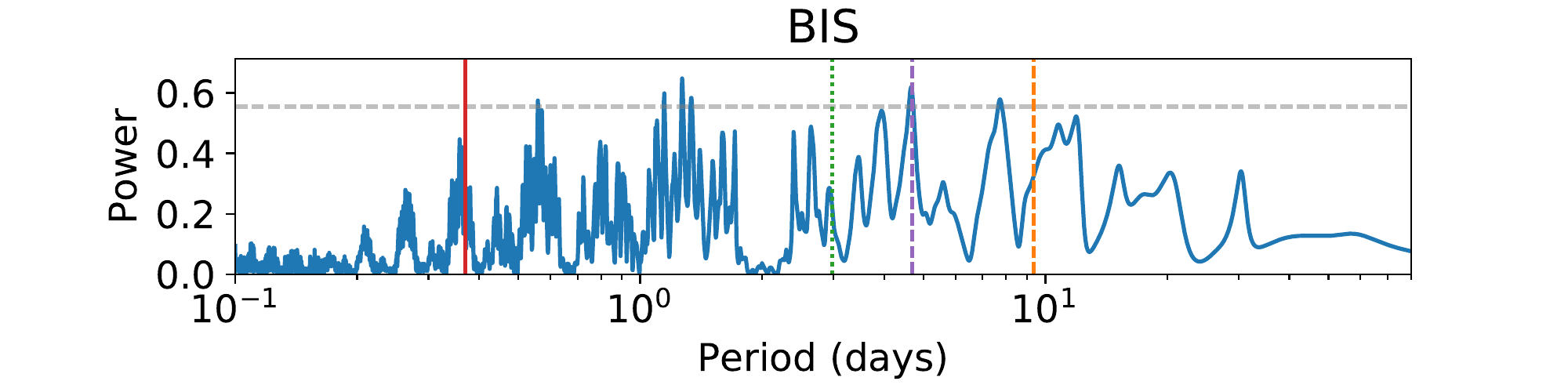}
\includegraphics[width = 1.\columnwidth]{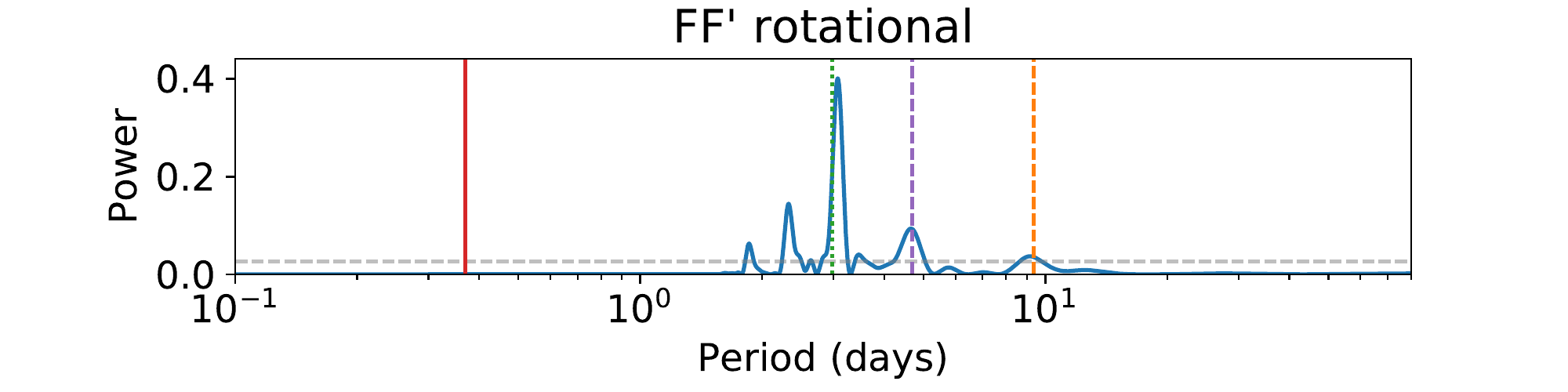}
\includegraphics[width = 1.\columnwidth]{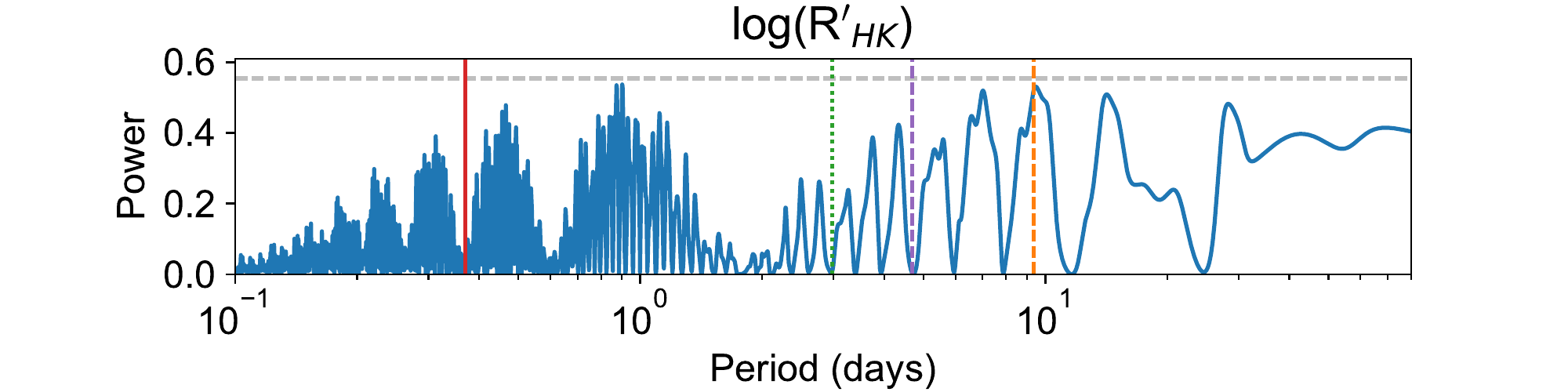}
\includegraphics[width = 1.\columnwidth]{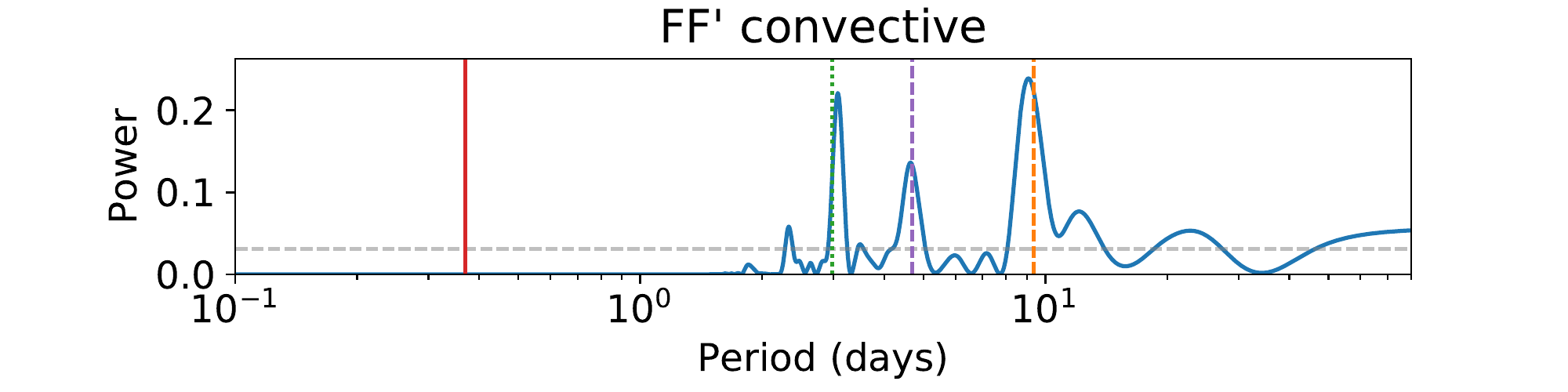}
\includegraphics[width = 1.\columnwidth]{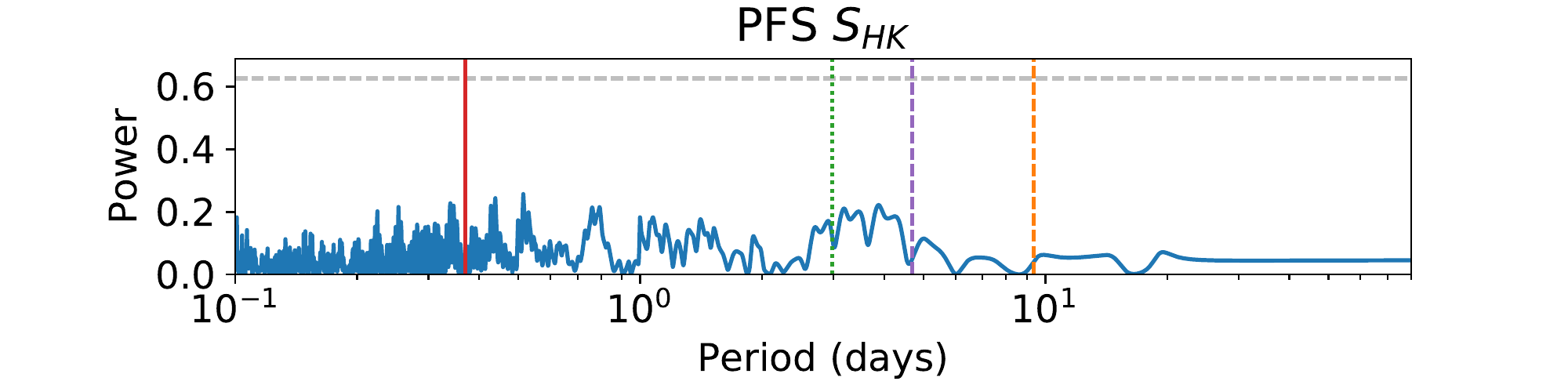}
\includegraphics[width = 1.\columnwidth]{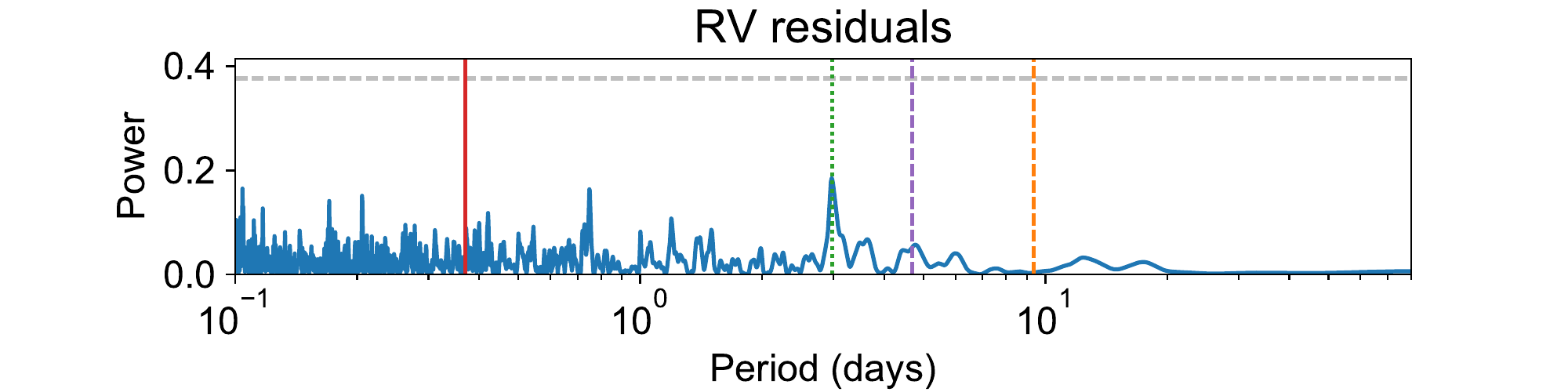}

\caption{Lomb-Scargle periodograms of the photometric data, RV data, and various activity indicators for EPIC~228732031. We also include the activity-induced RV variation predicted by $FF^\prime$ method \citep{Aigrain2012} and {\it K2} light curve. The red solid line is the orbital period of planet b. The orange dashed line is the stellar rotation period measured from the {\it K2} light curves and the purple dashed line is its first harmonic. The gray dashed line shows the power level for which the
false alarm probability is 0.001. The green dotted line is the strongest peak in the RV dataset near 3.0 days. Comparison with the activity-induced RV variation predicted by $FF^\prime$ method \citep{Aigrain2012} shows that 3.0-day periodicity is likely due to stellar activity.}
\label{periodograms}
\end{center}
\end{figure*}

Many of the detected USP planets have planetary companions (See Tab. \ref{usp}). Although {\it K2} light curve did not reveal another transiting planet (Section \ref{transit_detection}), there could be the
signal of a non-transiting
planet lurking in our radial velocity dataset. We addressed this question by scrutinizing the Lomb-Scargle periodograms of the light curves, radial velocities, and various activity indicators (See Fig. \ref{periodograms}).

The stellar rotation period of 9.4 days showed up clearly in the both the {\it K2} and the ground-based light curves. However the same periodicity is not significant in the periodogram of the measured
radial velocities. The strongest peak in the RV periodogram occurs at $3.0^{+0.19}_{-0.06}$ days with a
false-alarm probability (FAP) $<$0.001, shown with a green dotted line in Fig.~\ref{periodograms}.
This raises the question of whether this 3-day signal is due to a non-transiting planet,
or stellar activity. Based on the following reasons, we argue that stellar activity is the more likely
possibility.

The signal at $3.0^{+0.19}_{-0.06}$ days is suspiciously close the second harmonic $P_{\text{rot}}$/3 of the stellar rotation period (9.4 days). Previous simulations by \citet{Vanderburg2016RV} showed that the radial velocity variations induced by stellar activity often have a dominant periodicity at the first or second harmonics of the stellar rotation period (see their Fig. 6). \citet{Aigrain2012} presented the $FF^\prime$ method as a simple way to predict the radial velocity variations induced by stellar activity using precise and well-sampled light curves. Using the prescription provided by \citet{Aigrain2012} and the {\it K2} light curve, we estimated the activity-induced radial velocity variation of EPIC~228732031. As noted at the beginning of Section \ref{rv_analysis}, the activity-induced radial velocity variation has both rotational and convective components, which are represented
by Eqns.\ 10 and 12 of \citet{Aigrain2012}. For EPIC~228732031, the Lomb-Scargle periodograms of both the rotational and convective components showed a strong periodicity at $P_{\text{rot}}$/3 (see Fig.~\ref{periodograms}). This suggests that the $3.0^{+0.19}_{-0.06}$ days periodicity in the measured RV is attributable to activity-induced RV variation rather than a non-transiting planet.

\section{Discussion}

\subsection{Composition of EPIC~228732031b}

\begin{figure}[!t]
\begin{center}
\includegraphics[width = .8\columnwidth]{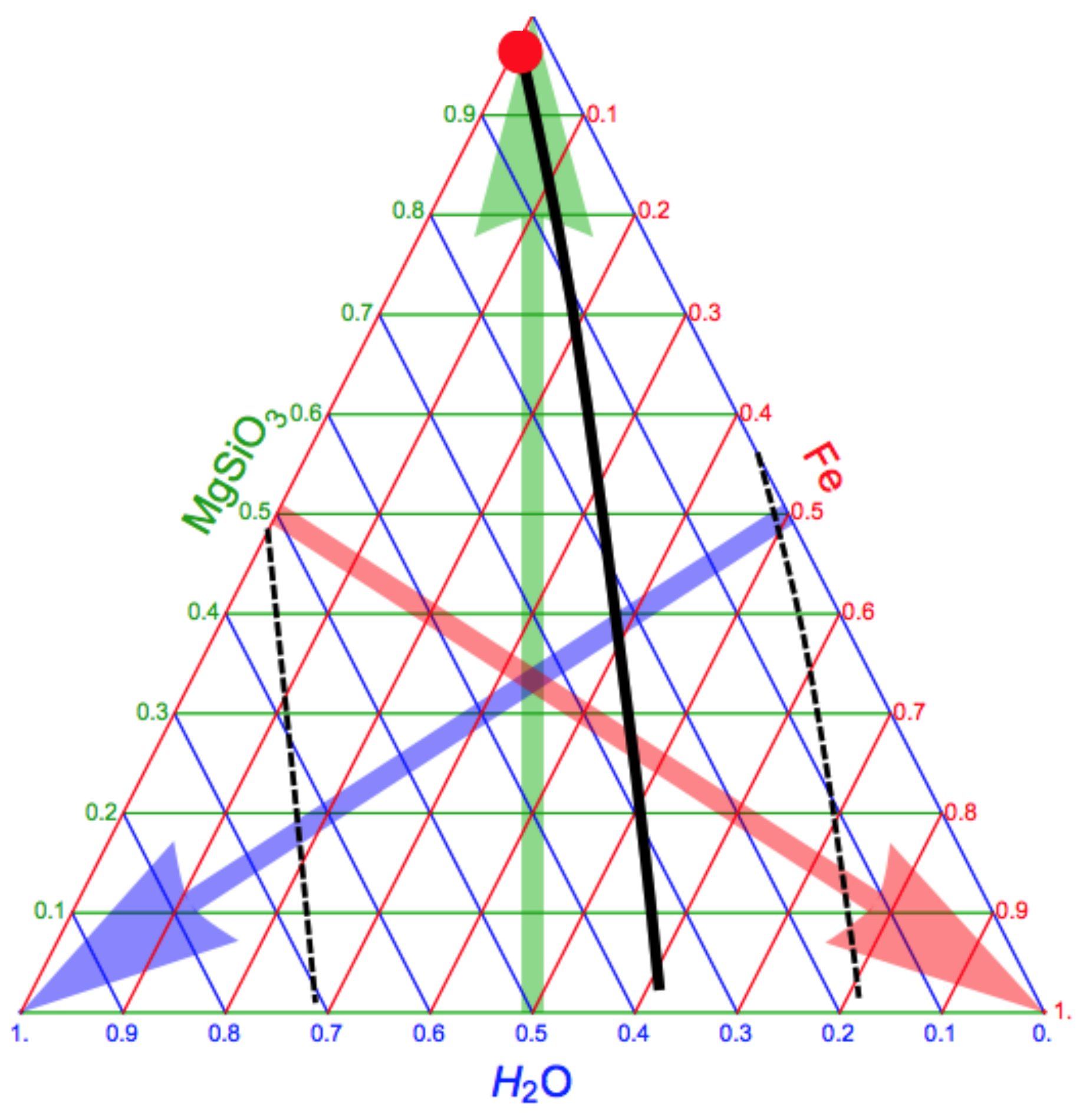}
\caption{The H$_2$O-MgSiO$_3$-Fe ternary diagram for the interior composition of EPIC~228732031b \citep{Zeng2016}. The solid line is the locus corresponding to the best-fit values of the planetary radius and mass. The dashed lines show the 1$\sigma$ confidence interval.}
\label{ternary}
\end{center}
\end{figure}

To investigate the constraints on the composition of  EPIC~228732031b,
we appeal to the theoretical models of the interiors of super-Earths by \citet{Zeng2016}.
We initially considered a differentiated three-component model
consisting of water, iron, and rock (magnesium silicate). We found the constraints on composition
to be quite weak. For EPIC~228732031b, the 1$\sigma$ confidence interval encompasses most of the iron/water/rock ternary diagram (see Fig. \ref{ternary}).

We then considered two more restricted models. In the first model, the planet is a mixture of rock and iron only, without the water component. For EPIC~228732031b, the iron fraction can be 0-44\% and still satisfy the 1$\sigma$ constraint on the planetary mass and radius.

The second model retains all three components --- rock, iron, and water --- but requires the iron/rock ratio to be $3/7$, similar
to the Earth. Thus in this model we determine the allowed range for the water component.
For EPIC~228732031b, the allowed range for the water mass fraction is 0-59\%.%

\subsection{Composition of the sample of USP planets}

As we saw in the preceding section, the measurement of mass and radius alone does not place
strong constraints on the composition of an individual planet. In the super-Earth regime (1-2 $R_{\oplus}$ or 1-10 $M_{\oplus}$),
there are many plausible compositions, which are difficult to distinguish based only on the mass and radius. To pin down the composition of any individual system, it will be necessary
to increase the measurement precision substantially or obtain additional information, such as
measurements of the atmospheric composition. With 8 members, though, the sample of USP planets has now grown to the point at which trends in composition with size, or other parameters, might start to become apparent.

\begin{figure*}
\begin{center}
\includegraphics[width = 1.5\columnwidth]{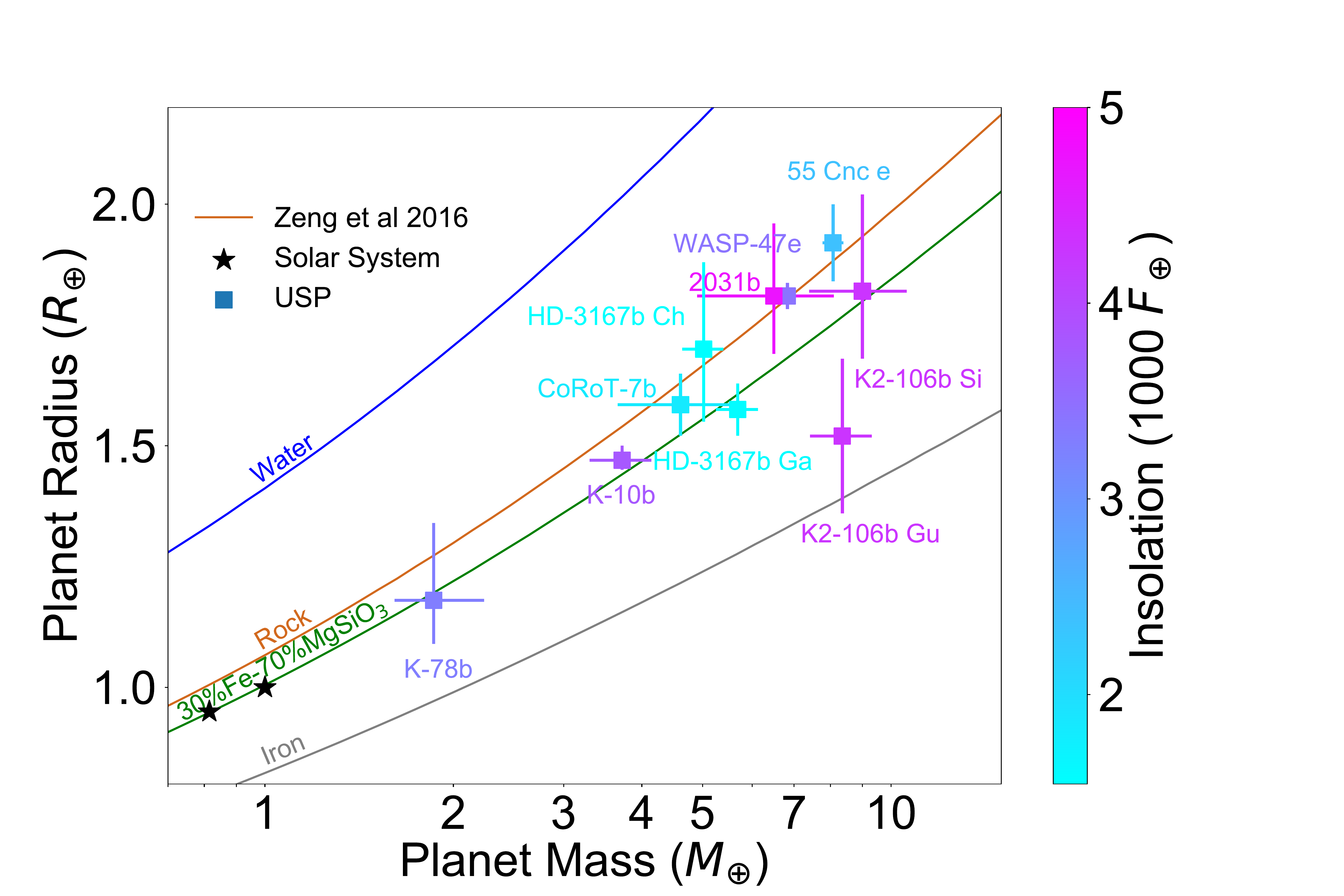}
\caption{The mass and radius of USPs along with theoretical mass-radius curves from \citet{Zeng2016}. The stars are solar system planets. Color indicates the level of insolation in units of $F_{\oplus}$, the insolation level received by Earth. For K2-106b, we plot the results from both \citet{Guenther2017} and \citet{Sinukoff2017}. For HD 3167b, we plot the results from both \citet{Christiansen2017} and \citet{Gandolfi2017}.}
\label{rm}
\end{center}
\end{figure*}

Table \ref{usp} summarizes the properties of all the known USP planets for which both
the planetary mass and radius have been measured. The table has been arranged in order of increasing planetary radius.
Figure~\ref{rm} displays their masses and radii, along with representative theoretical
mass-radius relationships from \citet{Zeng2016}. We did not include KOI-1843.03 \citep{Rappaport2013} and EPIC~228813918b \citep{Smith2017} in this diagram since their masses have not been measured, although
in both cases a lower limit on the density can be
obtained by assuming the planets are outside of
the Roche limit. The data points
are color-coded according to the level of irradiation by the star. One might have expected the more strongly irradiated planets to have a higher density,
as a consequence of photoevaporation.
However, we do not observe any correlation between planetary mean density and level of irradiation.
This may be because all of the USP planets are so strongly irradiated that photoevaporation has gone
to completion in all cases.
\citet{Lundkvist2016} argued for a threshold of $650~F_{\oplus}$  (where $F_{\oplus}$ is the insolation level received by Earth) as the value above
which close-in sub-Neptunes have undergone photoevaporation.
All of the USP planets in Table \ref{usp} have much higher levels of irradiation than this threshold.
Therefore, it is possible that all these planets have been entirely stripped of any pre-existing hydrogen/helium atmospheres,
and additional increases in irradiation would not affect the planetary mass or radius.
\citet{Ballard2014} also found no correlation between irradiation and mean density within a sample
of planets with measured masses, radii smaller than $2.2~R_{\oplus}$, and orbital periods $\lesssim$ 10 days.

In the mass-radius diagram (Fig. \ref{rm}), the 8 USP planets cluster between the theoretical relations
for pure rock (100\%~MgSiO$_3$) and an Earth-like composition (30\%~Fe and 70\%~MgSiO$_3$).
Earlier work by \citet{Dressing2015} pointed out that the best-characterized planets with masses smaller
than 6$M_{\oplus}$ are consistent with a composition of 17\%~Fe and 83\%~MgSiO$_3$.
Their sample of planets consisted of Kepler-78b, Kepler-10b, CoRoT-7b, Kepler-93b, and Kepler-36b.
Of these, the first 3 are USP planets; the latter two have orbital periods of 4.7 and 13.8~days.
\citet{Dressing2015} also claimed that planets heavier than $6~M_{\oplus}$ usually have a gaseous H/He envelope and/or a significant contribution of low-density
volatiles---presumably water---to the planet's total mass. Similarly \citet{Rogers2015} sought evidence for a critical planet radius separating
rocky planets and those with gaseous or water envelopes. She found that for planets with orbital periods shorter than 50 days, those that are smaller than $1.6~R_{\oplus}$ are predominantly rocky, whereas larger planets usually have a gaseous or volatile-enhanced envelopes.

Could the same hold true for the USP planets?
That is, could the larger and more massive USP planets hold on to a substantial water envelope, despite the extreme irradiation they experience?
\citet{Lopez2016} investigated this question from a theoretical point of view and argued that such an envelope could withstand photoevaporation.  He found that USP planets are able to retain water
envelopes even at irradiation levels of about 2800~$F_{\oplus}$.  If this is true, then we would
expect the USP planets larger than
1.6~$R{\oplus}$ to be more massive than $6~M_\oplus$ and
to have densities low enough to be compatible
with a water envelope.

Of the eight USP planets for which mass and radius have both been measured, five are
larger than $1.6~R_{\oplus}$: 55~Cnc~e, WASP-47e, EPIC~228732031b, HD~3167b and K2-106b. Interestingly, they also have masses heavier than $6~M_{\oplus}$. The first three of these (55~Cnc~e, WASP-47e, EPIC~228732031b) have a low mean density compatible with a
water envelope. Applying the second model described above, in which a "terrestrial" core (30\% iron, 70\% rock) is supplemented with
a water envelope, we find that these four planets all have water mass fractions $>$10\% at the best-fit values of their mass and radius. For K2-106b and HD~3167b,
two different groups have reported
different values for the masses and radii, leading to different conclusions about their composition. For K2-106b, \citet{Sinukoff2017} reported a planetary mass and radius of $9.0 \pm 1.6~M_{\oplus}$ and 1.82$^{+0.20}_{-0.14}~R_{\oplus}$ suggesting a rocky composition. \citet{Guenther2017} reported a planetary mass and radius of $8.36^{0.96}_{-0.94}~M_{\oplus}$ and $1.52 \pm 0.16 ~R_{\oplus}$ which pointed to an iron-rich composition. In both cases, the mean density of K2-106b seems to defy the simple interpretation that planets
more massive than $6~M_\oplus$ have H/He or water
envelopes. For HD~3167b, \citet{Christiansen2017} reported a planetary mass and radius of $5.02 \pm 0.38~M_{\oplus}$ and 1.70$^{+0.18}_{-0.15}~R_{\oplus}$ (consistent with water mass fractions of $>$10\%), whereas \citet{Gandolfi2017} reported $5.69 \pm 0.44~M_{\oplus}$ and 1.575$\pm 0.054~R_{\oplus}$, suggesting a predominantly rocky composition. Given the different results reported by the various groups and the fact that the radii of K2-106b and HD~3167b lie so close to the transition radius of $1.6~R_{\oplus}$ identified by \citet{Rogers2015}, the interpretation of these two planets is still unclear. More data, or at least a joint analysis of all the data
collected by both groups,
would probably help to clarify the situation.

A substantial water envelope
for 55 Cnc e, WASP-47e and EPIC~228732031b would have implications for
the formation of those planets. Theories in which these planets form {\it in situ} (i.e., near their
current orbits) would have difficulty explaining the presence of icy material
and the runaway accretion of the gas giant within the snow line.  Thus a massive water envelope
would seem to imply that the planet formed
beyond the snow line, unless there were some efficient mechanism for delivering
water to the inner portion of the planetary system.
To move a planet from beyond
the snow line to a very tight orbit, theorists have invoked disk migration, or
high-eccentricity migration. The observed architectures of the USP planetary systems, described below,
seem to be dynamically cold, and therefore more compatible
with disk migration than high-eccentricity migration.

It is interesting to note
that 55 Cnc e and WASP-47b, the most extensively studied planets in the USP planet sample, show some striking similarities.
Both systems appear to be dynamically cold.
The inner three planets of WASP-47 are all transiting, indicating low mutual inclinations \citep{Becker2015}, and \citet{Sanchis-Ojeda2015} ruled out a very large stellar obliquity ($\lambda$ = 0 $\pm$ 24$^{\circ}$). Dynamical analysis of the 55 Cnc system suggests that the
system is nearly coplanar, which helps maintain long-term stability \citep{Nelson2014}.
The dynamically cold configuration of both 55 Cnc and WASP-47 seems to disfavor any formation
scenario involving high-eccentricity migration, which would likely produce a dynamically hot final state. These two systems also share a few other intriguing and potentially relevant properties.
Among the USP planet sample, they have the largest number of detected planets (5 and 4), and are the only
systems in which gas giants are also known to be present (in 14.6-day and 4.2-day orbits).
They also have the most metal-rich stars, with [Fe/H] of 0.31 $\pm$ 0.04 and 0.36 $\pm$ 0.05,
as compared to the mean [Fe/H] of 0.0018 $\pm$ 0.0051 for the 62 USP systems studied by \citet{Winn2017}. The unusually high metallicity of these two systems might help to explain the large number of detected planets and the presence of giant planets in these systems. As argued by \citet{Dawson2016}, a metal-rich protoplanetary disk likely has a higher solid surface density. The higher solid surface density facilitates the formation and assembly of planet embryos. As a result, we might expect
a metal-rich disk to spawn more planets, and to facilitate the growth of solid cores past the critical core mass needed for giant planet formation.

\subsection{Other properties of the USP planets}

\citet{Sanchis-Ojeda_usp} pointed out that USP planets commonly have wider-orbiting planetary companions.
About 10\% of their sample of USP planets had longer-period transiting companions. After taking into
account the decline in transit probability with orbital period,
they inferred that nearly all the planets in their sample are likely to have longer-period companions, most of which are not transiting.  Of the 8 USP planets in Table~\ref{usp}, six of them have confirmed wider-orbiting
companions. The only known exceptions are Kepler-78 and EPIC~228732031, which are
probably also the least explored for additional planets. A problem in both cases is the high level of stellar
activity:
the light curves display clear rotational modulations with periods of 12.5 days and 9.4 days,
and the existing RV data show evidence for activity-related noise that hinders
the detection of additional planets \citep{Sanchis-Ojedak78,Howard2013}.

In summary, observations of systems with USP planets have revealed the following properties:
(1) They tend to have a high multiplicity of planets.
(2) The high multiplicity and the measured
properties of a few USP systems suggest that they
have a dynamically cold architecture characterized by low mutual inclinations and (possibly) a low stellar obliquity.
(3) The metallicity distribution of USP planet hosts is inconsistent with the metallicity distribution of hot Jupiter hosts,
and indistinguishable from that of the close-in sub-Neptunes \citep{Winn2017}.
(4) At least three of the five USP planets
known to be larger than $1.6~R_{\oplus}$ are
also heavier than $6~M_{\oplus}$ and have densities
low enough to be compatible with a water envelope.
Taken together these observations suggest to us
that the USP planets are a subset of the general
population of sub-Neptune planets, most of which have lost their atmospheres entirely to photoevaporation, except
for the largest few that have retained a substantial water envelope.

\acknowledgments
This work was carried out as part of the KESPRINT consortium. This paper includes data gathered with the 6.5 meter Magellan Telescopes located at Las Campanas Observatory, Chile. Based on observations made with the Italian Telescopio Nazionale Galileo (TNG) operated on the island of La Palma by the Fundaci\'on Galileo Galilei of the INAF (Istituto Nazionale di Astrofisica) at the Spanish Observatorio del Roque de los Muchachos of the Instituto de Astrof\'isica de Canarias, as part of the observing programs A34TAC\_10 and A34TAC\_44. We are very grateful to the TNG staff members for their precious support during the observations. Some of the data presented in this work were obtained at the WIYN Observatory from telescope time allocated to NN-EXPLORE through the scientific partnership of the National Aeronautics and Space Administration, the National Science Foundation, and the National Optical Astronomy Observatory. This work was supported by a NASA WIYN PI Data Award, administered by the NASA Exoplanet Science Institute. NESSI was funded by the NASA Exoplanet Exploration Program and the NASA Ames Research Center. NESSI was built at the Ames Research Center by Steve B. Howell, Nic Scott, Elliott P. Horch, and Emmett Quigley. The authors wish to recognize and acknowledge the very significant cultural role and reverence that the summit of Maunakea has always had within the indigenous Hawaiian community. We are most fortunate to have the opportunity to conduct observations from this mountain. The authors are honored to be permitted to conduct observations on Iolkam Du’ag (Kitt Peak), a mountain within the Tohono O'odham Nation with particular significance to the Tohono O'odham people. D.\,G. gratefully acknowledges the financial support of the \emph{Programma Giovani Ricercatori -- Rita Levi Montalcini -- Rientro dei Cervelli (2012)} awarded by the Italian Ministry of Education, Universities and Research (MIUR).

\facilities{{\it  Kepler (K2)}, Fairborn Observatory/AIT, WIYN/NESSI, LCO/Swope, Magellan/PFS, TNG/HARPS-N} 
\software{{\tt ATLAS12} \citep{Kurucz2013}, {\tt Python}, {\tt Batman}\citep{Kreidberg2015}, {\tt emcee} \citep{emcee}, {\tt iSpec} \citep{Blanco-Cuaresma2014}, {\tt Isochrones}\citep{Morton2015}, {\tt lmfit} \citep{lmfit}, {\tt MARCS} \citep{Gustafsson2008}, {\tt MOOG} \citep{Sneden1973}, {\tt MultiNest}\citep{Feroz2009}}, {\tt NEXTGEN} \citep{Hauschildt1999},  {\tt scipy} \citep{Jones2001}, {\tt SME} \citep[v5.22][]{vp96,vf05,pv2017}, {\tt SpecMatch-emp6}\citep{Yee2017}, {\tt SPECTRUM3} \citep[V2.76][]{Gray1994}, {\tt VALD3}\citep{pk1995,kp1999}

\bibliographystyle{yahapj}
\bibliography{main}

\startlongtable
\begin{deluxetable*}{ccc}
\tablecaption{AIT Photometry \label{tab:AIT}}
\tablehead{
\colhead{BJD$_{\text{TDB}}$} & \colhead{$\Delta$ R} & \colhead{Unc}
}
\startdata
2457827.69557  &  -0.92694  &  0.00264 \\
2457827.73587  &  -0.9255  &  0.00707 \\
2457827.80667  &  -0.93827  &  0.00153 \\
2457827.85367  &  -0.93305  &  0.00046 \\
2457827.89417  &  -0.93372  &  0.00144 \\
2457827.93727  &  -0.93393  &  0.00145 \\
...
\enddata
\end{deluxetable*}

\startlongtable
\begin{deluxetable*}{ccc}
\tablecaption{Swope Photometry \label{tab:Swope}}
\tablehead{
\colhead{BJD$_{\text{TDB}}$} & \colhead{Relative Flux} & \colhead{Unc.}}
\startdata
2457834.08327 &  0.9965 & 0.0035 \\
2457834.35326 &  0.9972 & 0.0041 \\
2457835.10169 &  0.9979 & 0.0046 \\
...
\enddata
\end{deluxetable*}

\startlongtable

\begin{deluxetable*}{cccccccc}

\tablecaption{HARPS-N Observations \label{HARPS-N}}
\tablehead{
\colhead{BJD$_{\text{TDB}}$} & \colhead{RV (m s$^{-1}$)} & \colhead{Unc (m s$^{-1}$)} & \colhead{FWHM}  & \colhead{BIS} & \colhead{log\,$R^\prime_\mathrm{HK}$} & \colhead{$\Delta$ log\,$R^\prime_\mathrm{HK}$} & \colhead{SNR}
}
\startdata
2457782.65615  &  -6682.78  &  3.71  &  7.858  &  0.0465  &  -4.512  &  0.013  &  39.8 \\
2457783.61632  &  -6710.55  &  5.95  &  7.870  &  0.0412  &  -4.537  &  0.027  &  26.1 \\
2457783.72195  &  -6698.96  &  8.76  &  7.827  &  0.0315  &  -4.471  &  0.040  &  18.5 \\
2457812.59115  &  -6672.32  &  4.00  &  7.807  &  -0.0042  &  -4.519  &  0.015  &  35.5 \\
2457812.66810  &  -6672.00  &  4.22  &  7.814  &  0.0146  &  -4.536  &  0.016  &  35.9 \\
2457812.72111  &  -6690.31  &  4.63  &  7.820  &  0.0212  &  -4.538  &  0.019  &  34.2 \\
2457813.53461  &  -6701.27  &  4.12  &  7.850  &  0.0486  &  -4.534  &  0.016  &  36.6 \\
2457813.56281  &  -6700.71  &  5.34  &  7.869  &  0.0139  &  -4.566  &  0.024  &  29.3 \\
2457813.58430  &  -6695.97  &  3.67  &  7.864  &  0.0398  &  -4.524  &  0.013  &  40.0 \\
2457813.60761  &  -6705.87  &  4.23  &  7.859  &  0.0297  &  -4.533  &  0.016  &  35.2 \\
2457813.63498  &  -6694.27  &  5.12  &  7.848  &  0.0516  &  -4.509  &  0.019  &  29.9 \\
2457813.65657  &  -6699.99  &  5.00  &  7.859  &  0.0445  &  -4.490  &  0.018  &  30.3 \\
2457813.68168  &  -6696.17  &  10.43  &  7.849  &  -0.0087  &  -4.426  &  0.041  &  16.7 \\
2457836.48220  &  -6675.40  &  7.03  &  7.943  &  0.0431  &  -4.496  &  0.030  &  22.3 \\
2457836.50805  &  -6665.92  &  7.48  &  7.932  &  0.0124  &  -4.489  &  0.032  &  21.4 \\
2457836.53315  &  -6661.90  &  6.03  &  7.951  &  0.0321  &  -4.490  &  0.024  &  25.1 \\
2457836.55735  &  -6657.92  &  5.08  &  7.914  &  0.0369  &  -4.480  &  0.018  &  29.4 \\
2457836.58220  &  -6667.00  &  4.72  &  7.925  &  0.0032  &  -4.470  &  0.016  &  31.8 \\
2457836.60656  &  -6668.85  &  3.97  &  7.919  &  0.0114  &  -4.470  &  0.012  &  33.4 \\
2457836.63080  &  -6671.00  &  4.38  &  7.927  &  0.0148  &  -4.500  &  0.015  &  31.5 \\
2457836.65503  &  -6671.76  &  5.12  &  7.929  &  0.0235  &  -4.497  &  0.020  &  28.5 \\
2457837.47474  &  -6707.18  &  5.09  &  7.967  &  0.0404  &  -4.473  &  0.018  &  30.4 \\
2457837.49701  &  -6709.14  &  6.08  &  7.906  &  0.0589  &  -4.499  &  0.025  &  26.0 \\
2457837.51966  &  -6714.79  &  4.02  &  7.905  &  0.0667  &  -4.493  &  0.013  &  36.9 \\
2457837.54103  &  -6704.63  &  5.38  &  7.914  &  0.0446  &  -4.458  &  0.019  &  28.8 \\
2457837.56389  &  -6712.95  &  12.03  &  7.921  &  0.0510  &  -4.483  &  0.058  &  14.8 \\
2457837.58317\tablenotemark{*}  &  -6892.80  &  77.70  &  7.892  &  0.1897  &  -4.139  &  0.238  &  2.7 \\
2457838.52396  &  -6720.53  &  3.27  &  7.839  &  0.0669  &  -4.510  &  0.010  &  40.6 \\
2457838.54842  &  -6726.57  &  3.42  &  7.852  &  0.0770  &  -4.491  &  0.010  &  40.3 \\
2457838.57294  &  -6719.73  &  3.34  &  7.835  &  0.0722  &  -4.502  &  0.010  &  40.7 \\
2457838.59768  &  -6725.33  &  3.72  &  7.842  &  0.0700  &  -4.505  &  0.012  &  36.9 \\
2457838.62169  &  -6712.55  &  4.64  &  7.828  &  0.0646  &  -4.488  &  0.017  &  31.6 \\
2457838.64668  &  -6713.49  &  4.98  &  7.838  &  0.0550  &  -4.530  &  0.021  &  29.9 \\
2457839.49799\tablenotemark{*}  &  -6670.48  &  106.24  &  7.765  &  0.3354  &  -4.382  &  0.538  &  1.5 \\
2457844.44051  &  -6700.01  &  4.63  &  7.884  &  0.0233  &  -4.484  &  0.017  &  32.4 \\
2457844.47029  &  -6700.62  &  4.24  &  7.854  &  0.0237  &  -4.494  &  0.015  &  35.4 \\
2457844.49442  &  -6697.37  &  4.55  &  7.855  &  0.0295  &  -4.497  &  0.017  &  32.3 \\
2457844.51857  &  -6701.16  &  5.18  &  7.851  &  0.0117  &  -4.507  &  0.021  &  30.0 \\
2457844.54365  &  -6699.62  &  6.26  &  7.855  &  0.0452  &  -4.473  &  0.025  &  25.1 \\
2457844.56786  &  -6695.05  &  6.64  &  7.865  &  0.0292  &  -4.513  &  0.030  &  23.7 \\
2457844.59224  &  -6688.51  &  8.83  &  7.877  &  0.0147  &  -4.485  &  0.041  &  19.3 \\
\enddata
\tablenotetext{*}{Excluded from analysis due to bad seeing.}
\end{deluxetable*}

\clearpage

\startlongtable

\begin{deluxetable*}{cccc}
\tablecaption{PFS Observations \label{PFS}}
\tablehead{
\colhead{BJD$_{\text{TDB}}$} & \colhead{RV (m s$^{-1}$)} & \colhead{Unc (m s$^{-1}$)} & \colhead{S$_{\text{HK}}$}
}
\startdata
2457828.85718  &  -45.17  &  3.88  &  0.701 \\
2457828.87266  &  -36.47  &  4.03  &  0.658 \\
2457829.70652  &  -20.52  &  3.55  &  0.534 \\
2457829.72141  &  -4.58  &  6.07  &  0.618 \\
2457830.59287  &  17.71  &  3.56  &  0.552 \\
2457830.60890  &  23.34  &  3.48  &  0.492 \\
2457830.68486  &  21.34  &  3.51  &  0.478 \\
2457830.70112  &  19.92  &  3.26  &  0.498 \\
2457830.74201  &  -0.74  &  3.44  &  0.474 \\
2457830.75830  &  16.90  &  3.50  &  0.489 \\
2457830.83794  &  8.25  &  4.64  &  0.503 \\
2457830.85420  &  -4.06  &  4.91  &  0.803 \\
2457832.60032  &  -29.51  &  4.96  &  0.635 \\
2457832.61610  &  -21.56  &  4.11  &  0.603 \\
2457832.69542  &  -25.76  &  5.59  &  0.693 \\
2457832.71105\tablenotemark{*}  &  -51.72  &  10.95  &  0.742 \\
2457833.62080  &  2.10  &  3.91  &  0.596 \\
2457833.63700  &  4.38  &  3.73  &  0.595 \\
2457833.71138  &  2.23  &  4.12  &  0.527 \\
2457833.72778  &  -1.98  &  4.06  &  0.539 \\
2457833.82991  &  0.55  &  3.54  &  0.502 \\
2457833.84582  &  -4.49  &  3.69  &  0.560 \\
2457848.53428  &  9.40  &  4.49  &  0.668 \\
2457848.55023  &  2.09  &  4.17  &  0.658 \\
2457848.63414  &  7.27  &  4.23  &  0.687 \\
2457848.65008  &  6.40  &  4.15  &  0.562 \\
2457848.70933  &  29.22  &  3.92  &  0.492 \\
2457848.72551  &  14.37  &  4.19  &  0.572 \\
2457848.75682  &  23.83  &  5.51  &  0.616 \\
2457848.77330  &  6.85  &  5.44  &  0.562 \\
2457848.79458  &  19.56  &  5.14  &  0.437 \\
2457848.81068  &  10.63  &  5.45  &  0.885 \\
\enddata
\tablenotetext{*}{Excluded from analysis due to bad seeing.}
\end{deluxetable*}
\clearpage

\begin{deluxetable*}{cccccc}
\tablecaption{Spectroscopic Parameters of EPIC~228732031 \label{spectro}}
\tablehead{
\colhead{Parameters} & \colhead{Method 1} & \colhead{Method 2} & \colhead{Method 3}& \colhead{Method 4}& \colhead{Adopted}
}
\startdata
$T_{\text{eff}} ~(K)$ &$ 5225\pm 70     $&$ 5216\pm 27     $ &$ 4975\pm 125     $&$ 5100\pm 110     $&$ 5200\pm 100     $\\
$\logg~(\text{dex})$ &$ 	4.67\pm 0.08  $&$ 4.63\pm 0.05     $&$ 4.40\pm 0.15     $&$ 4.60\pm 0.10     $&$ 4.62\pm 0.10     $\\
$[\text{Fe/H}]~(\text{dex})$ &$ 	0.01\pm 0.05  $&$ -0.02\pm 0.09     $&$ -0.06\pm 0.10     $&$ -0.06\pm 0.09     $&$ -0.02\pm 0.08     $ \\
\vsini (km~s$^{-1}$) &$ 	4.8\pm 0.6  $&$ 4.0\pm 0.6     $&$ 4.8\pm 1.6     $& &$ 4.4\pm 1.0     $\\
\enddata

\end{deluxetable*}

\begin{deluxetable*}{ccc}
\tablecaption{Stellar Parameters of EPIC~228732031 \label{stellar}}
\tablehead{
\colhead{Parameters} & \colhead{Value and 68.3\% Conf. Limits} & \colhead{Ref.}}

\startdata
R.A. ($^{\circ}$) & 182.751556 & A\\
Dec. ($^{\circ}$) & -9.765218 & A\\
V (mag) & 12.115 $\pm$ 0.020& A\\
$T_{\text{eff}} ~(K)$ &$ 5200\pm 100     $& B \\
$\log~g~(\text{dex})$ &$ 	4.62\pm 0.10  $& B \\
$[\text{Fe/H}]~(\text{dex})$ &$ 	-0.00\pm 0.08  $& B \\
$v~\text{sin}~i$ ~(km~s$^{-1})$ &$ 4.4 \pm 1.0     $& B \\
$M_{\star} ~(~M_{\odot})$ &$ 0.84 \pm 0.03     $& B \\
$R_{\star} ~(R_{\odot})$ &$ 0.81 \pm 0.03     $& B \\
$P_{\text{rot}}$ (days) &$ 9.37 \pm 1.85     $&B \\
$\rho_{\text{spe}}$ (g cm$^{-3}$)\tablenotemark{1} &$ 2.23 \pm 0.33     $&B \\
$\rho_{\text{tra}}$ (g cm$^{-3}$)\tablenotemark{2} &$ 2.43 ^{+0.61}_{-1.09}$   &B \\
u$_1$ & 0.53$\pm$ 0.10& B\\
u$_2$ & 0.20$\pm$ 0.10& B\\
$A_\mathrm{v}$ (mag)& $0.07\pm0.05$&B\\
$d$ (pc)&$174\pm20$&B\\
\enddata

\tablenotetext{1}{Mean density from the derived mass and radius.}
\tablenotetext{2}{Mean density from modeling the transit light curve.}
\tablecomments{A:ExoFOP; B: this work.}
\end{deluxetable*}

\begin{deluxetable*}{cc}
\tablecaption{Planetary Parameters of EPIC~228732031b \label{planet}}
\tablehead{
\colhead{Parameters} & \colhead{Value and 68.3\% Conf. Limits}}

\startdata
Transits &\\
$P_{\text{orb}} ~(days)$ &$ 0.3693038\pm 0.0000091     $\\
$t_{\text{c}}$ (BJD) &$ 2457582.9360 \pm 0.0011     $\\
$R_p/R_{\star}$ & 0.0204$^{+0.0010}_{-0.0006}$\\
$R_p~(R_{\oplus})$ & 1.81$^{+0.16}_{-0.12}$\\
$a/R_{\star}$ & 2.66$^{+0.18}_{-0.36}$\\
$i~(^{\circ})$ &85$^{+9}_{-10}$ \\
\hline
Floating Chunk Method &\\
$K$ (m s$^{-1})$ & $6.77 \pm 1.50$ \\
$M_p~(M_{\oplus})$  & 6.8$\pm$ 1.6\\
$\rho_p$~(g~cm$^{-3})$ & 6.3$^{+3.1}_{-2.8}$ \\
$\sigma_{jit, PFS}$ (m s$^{-1}$) & $5.3^{+1.6}_{-1.2}$ \\
$\sigma_{jit, HARPS-N}$ (m s$^{-1}$)& $2.0^{+1.6}_{-1.3}$ \\
$e$ & <0.26 (95\% Conf. Level) \\
\hline
Gaussian Process&\\
$h_{\text{rv}}$ (m s$^{-1}$)& $26.0^{+7.3}_{-5.1}$\\
$\tau$ (days)& $8.9 \pm 1.6$\\
$\Gamma$ & $4.18 \pm 0.94$\\
$T$ (days) & $9.68 \pm 0.15$\\
$\gamma_{\text{HARPS-N}}$ (m s$^{-1}$)& $-6694.7^{+12.1}_{-10.8}$\\
$\gamma_{\text{PFS}}$ (m s$^{-1}$)& $-1.0^{+11.7}_{-10.6}$\\
$\sigma_{\text{jit, HARPS-N}}$ (m s$^{-1}$)& $2.0^{+1.6}_{-1.3}$\\
$\sigma_{\text{jit, PFS}}$ (m s$^{-1}$)& $5.3^{+1.6}_{-1.2}$\\
$K$ (m s$^{-1})$ & $6.55 \pm 1.48$ \\
$e$&0(fixed)\\
$M_p~(M_{\oplus})$ &6.5$\pm 1.6$\\
$\rho_p$~(g~cm$^{-3})$ & 6.0$^{+3.0}_{-2.7}$ \\
\enddata

\end{deluxetable*}

\begin{deluxetable*}{lcccccccccccc}[h!]
\rotate
\fontsize{8}{7}
\tablecaption{Ultra-short-period planets with mass measurements\label{usp}}
\tablehead{\colhead{}  & \colhead{$T_{\text{eff}}$}  &\colhead{[Fe/H]} &\colhead{$M_\star$} & \colhead{$R_\star$}  & \colhead{P} & \colhead{$R_p$} & \colhead{$M_p$} & \colhead{$\rho_p$}  & \colhead{$N_{\rm{pl}}$}& \colhead{Fe-MgSiO$_3$\tablenotemark{1}} &\colhead{H$_2$O\tablenotemark{2}}& \colhead{Ref.} \\
 & \colhead{(K)}& \colhead{(dex)} & \colhead{($M_\odot$)} & \colhead{($R_\odot$)}  & \colhead{(days)} & \colhead{($R_\oplus$)} & \colhead{($M_\oplus$)} & \colhead{(g~cm$^{-3}$)}  & &  & }
\startdata
Kepler-78b & 5089 $\pm$ 50 & -0.14  $\pm$ 0.08 & 0.83 $\pm$ 0.05 & 0.74 $\pm$ 0.05 &  0.36 & 1.20$\pm$ 0.09 & 1.87$^{+0.27}_{-0.26}$ & 6.0$^{+1.9}_{-1.4}$ & 1 & 36\%-64\% & <57\%& Ho13, Gr15 \\
Kepler-10b & 5627 $\pm$ 44 & -0.09  $\pm$ 0.04 & 0.913 $\pm$ 0.022 & 1.065 $\pm$ 0.009 &  0.84 & 1.47$^{+0.03}_{-0.02}$& 3.72 $\pm$ 0.42 & 6.46 $\pm 0.73$ & 2 & 21\%-79\% & 2$^{+10}_{-2}$\% &  Ba11, We16\\
CoRot-7b & 5250 $\pm$ 60 & 0.12 $\pm$ 0.06 & 0.91 $\pm$ 0.03 & 0.82 $\pm$ 0.04 &  0.85 & 1.585 $\pm$ 0.064 & 4.73 $\pm$ 0.95 & 6.61 $\pm$ 1.33 & 2 &  15\%-85\%& 3$^{+23}_{-3}$\%&  Br10, Ha14 \\
K2-106b & 5470 $\pm$ 30 & 0.025 $\pm$ 0.020 & 0.945 $\pm$ 0.063 & 0.869 $\pm$ 0.088 &  0.57 & 1.52 $\pm$ 0.16 & 8.36$^{+0.96}_{-0.94}$ & 13.1$^{+5.4}_{-3.6}$ & 2 & 80\%-20\%& < 20\%  & Gu17 \\
K2-106b & 5496 $\pm$ 46 & 0.06 $\pm$ 0.03 & 0.95 $\pm$ 0.05 & 0.92 $\pm$ 0.03 &  0.57 & 1.82$^{+0.2}_{-0.14}$ & 9.0 $\pm$ 1.6 & 8.57$^{+4.64}_{-2.80}$ & 2 & 26\%-74\% & 2$^{+27}_{-2}$\%& Si17 \\
HD 3167b & 5286 $\pm$ 40 & 0.02 $\pm$ 0.03 & 0.877 $\pm$  0.024 & 0.835 $\pm$ 0.026 &  0.96 & 1.575 $\pm$ 0.054 & 5.69 $\pm$ 0.44 & 8.00$^{+1.10}_{-0.98}$ & 2-4 & 37\%-63\%& <10\% & Ga17 \\
HD 3167b & 5261 $\pm$ 60 & 0.04 $\pm$ 0.05 & 0.86 $\pm$  0.03 & 0.86 $\pm$ 0.04 &  0.96 & 1.70$^{+0.18}_{-0.15}$ & 5.02 $\pm$ 0.38 & 5.60$^{+2.15}_{-1.43}$ & 3 &  At least 6\% H$_2$O & 15$^{+35}_{-15}$\% & Ch17 \\
EPIC~228732031b & 5200 $\pm$ 100& 0.00 $\pm$ 0.08 & 0.84 $\pm$ 0.03 & 0.81 $\pm$ 0.03 &  0.37 & 1.81$^{+0.16}_{-0.12}$ & 6.5$\pm 1.6$ & 6.0$^{+3.0}_{-2.7}$ & 1 & At least 4\% H$_2$O &15$^{+44}_{-15}$\%& This Work\\
WASP-47e & 5576 $\pm$ 68 & 0.36 $\pm$ 0.05 & 1.04$\pm$  0.031 & 1.137 $\pm$ 0.013 &  0.79 & 1.81 $\pm$ 0.027 & 6.83 $\pm$ 0.66 &  6.35 $\pm$ 0.78  & 4 &  At least 2\% H$_2$O&12$^{+10}_{-8}$\% & Be15, Va17 \\
55 Cnc e & 5196 $\pm$ 24 & 0.31 $\pm$ 0.04 & 0.905 $\pm$ 0.015 & 0.943 $\pm$ 0.010  & 0.74 & 1.92 $\pm$ 0.08 & 8.08 $\pm$ 0.31 & 6.4$^{+0.8}_{-0.7}$ & 5 &  At least 4\% H$_2$O&16$^{+35}_{-10}$\% & Va05, Br11, De16\\
\enddata
\tablenotetext{1}{Composition from the differentiated, two-component (iron and magnesium silicate) model by \citet{Zeng2016}. The reported composition are calculated with the median values of the planet's mass and radius.}
\tablenotetext{2}{Water mass fraction assuming a water envelope on top of a Earth-like core of 30\%Fe-70\%MgSiO$_3$. The upper limits are quoted at 95\% confidence level.}
\tablecomments{Va05:\citet{Valenti2005}; Br11: \citet{Braun2011}; De16: \citet{Demory2016}; Br10: \citet{Bruntt2010}; Ha14: \citet{Haywood2014}; Si17: \citet{Sinukoff2017}; Gu17: \citet{Guenther2017}; Ga17: \citet{Gandolfi2017}; Ch17: \citet{Christiansen2017}; Ba11: \citet{Batalha2011}; We16: \citet{Weiss2016}; Ho13: \citet{Howard2013}; Gr15: \citet{Grunblatt2015}; Be15: \citet{Becker2015}; Va17: Vanderburg et al, submitted}
\end{deluxetable*}

\end{document}